\documentclass[
reprint,
onecolumn,
aps,
superscriptadress,
amsmath,amssymb,
notitlepage,
longbibliography]{revtex4-1}
\usepackage{graphicx,amsmath,amssymb,color}
\usepackage{dcolumn}% Align table columns on decimal point
\usepackage{subfigure}
\def\Rey{ \mathcal{R}e }
\usepackage{graphicx}
\usepackage{psfrag}
\usepackage{epstopdf}

\newcommand{\be}{\begin{equation}}
\newcommand{\ee}{\end{equation}}

\begin{document}

\title{Structure of the velocity gradient tensor 
in turbulent shear flows}

\author{Alain Pumir}
\affiliation{Laboratoire de Physique, Ecole Normale Sup\'erieure de Lyon, CNRS
and Universit\'e de Lyon, \\
46, all\'ee d'Italie, F-69007, France}

\date{\today}

\begin{abstract}
The expected universality of small-scale properties of turbulent 
flows implies isotropic properties of the velocity gradient tensor
in the very large Reynolds number limit. 
Using direct numerical simulations, we determine the tensors formed by 
$n = 2$ and $3$ velocity gradients at a single point in turbulent 
homogeneous shear flows, and in the log-layer of a turbulent channel flow,
and we characterize the departure of these tensors from the 
corresponding isotropic prediction.
Specifically, we separate the \textit{even} components of the tensors,
invariant under reflexion with respect to all axes, from the \textit{odd} ones,
which identically vanish in the absence of shear.
Our results indicate that the largest deviation from isotropy comes from 
the odd component of the third velocity gradient correlation function, 
especially from the third moment of the derivative along the normal
direction of the streamwise velocity component.
At the Reynolds numbers considered ($R_\lambda \approx 140$), we observe that
these second and third order correlation functions are significantly 
larger in turbulent channel flows than in homogeneous shear flow. 
Overall, our work demonstrates that a mean shear leads to relatively simple 
structure of the velocity gradient tensor. How isotropy is restored in the 
very large Reynolds limit  remains to be understood.
\end{abstract}

\maketitle

\section{Introduction }
\label{sec:intro}

In 3-dimensional turbulence, the rate of production of small scales by the 
flow, as well as the energy dissipation can be expressed in terms of
third and second order correlations of the
velocity gradient tensor $\mathbf{A}$
($A_{ab} \equiv \partial_a u_b$, where $\mathbf{u}$ is the fluid 
velocity)~\cite{Betchov56,Lumley}.
This remark provides a very strong motivation for investigating the 
statistical properties of $\mathbf{A}$. Here, we focus on the correlation
tensors obtained by averaging $n = 2$ or $n = 3$ velocity gradients taken
at a single point. These relatively low moments are not very
sensitive to the formation of very intense velocity gradients in high 
Reynolds turbulent flows~\cite{Batchelor+49,Frisch}, a phenomenon which is 
not discussed in this work.

The tensors obtained from $n = 2$ and $n = 3$ velocity gradients, measured
at a single point can be exactly determined in the simplified case of a 
homogeneous and 
isotropic turbulent flow~\cite{Betchov56,Siggia81}. The expected universality 
of the small-scale velocity fluctuations when the
Reynolds number, $\Rey$, is very large~\cite{K41} implies that the velocity 
tensor correlations should 
coincide with their isotropic forms in the $\Rey \rightarrow \infty$ limit. 
The aim of this work is to investigate, using direct numerical simulation 
(DNS) results,
the structure of the velocity gradient tensors in simple shear flows, 
namely in turbulent channel flows (TCF), and in turbulent homogeneous shear 
flows 
(HSF). Whereas the practical and fundamental interest in studying TCF is 
obvious~\cite{Lumley,Pope}, we stress that HSF 
provides an ideal setting to investigate the influence 
of a large scale shear on the small-scale properties of 
turbulence~\cite{Champagne:70,GargWarh98,PumShr95,Pumir+16}. Recent studies actually
point
to similarities between TCF and HSF, in terms of the mechanisms leading
to formation and development of large scale structures~\cite{Pumir96,Mizuno13,Dong16}.
Higher moments of the velocity gradient tensors have been investigated
in TCF, see e.g.~\cite{Vreman:14}. The third
moment of the derivative of the streamwise velocity component, in the direction
normal to the wall, $\partial_2 u_1$, has received special attention in 
relation with the 
issue of small-scale isotropy~\cite{PumShr95,Pumir96,GargWarh98,ShenWarh00}.
This quantity, which has been measured experimentally in HSF, points
to a slower than anticipated decay of anisotropy~\cite{ShenWarh00}.
This property is reminiscent of the strong and persistent anisotropy found
in the case of a passive scalar mixed by a turbulent flow in the presence
of a mean gradient. In this case, numerical and experimental results indicate
a skewness of the scalar gradient, parallel to the mean gradient, of order 
$1$, independent of the Reynolds number~\cite{Sreeni91,HolzSig94,Pumir94,TongWarh94,Warhaft00}.
A direct comparison between TCF and HSF, reveals similarities between the
properties of the two flows, although at comparable Reynolds numbers, $R_\lambda \approx 140$, it was found that the skewness of $\partial_2 u_1$ was 
roughly two times larger in the log-layer of the TCF than in HSF~\cite{Pumir+16}.

We investigate here the full second and third order correlations of 
the velocity gradient tensor in HSF and TCF, defined as:  
\begin{equation}
\overline{T}^{2,flow}_{abcd} = \langle \partial_a u_b \partial_c u_d \rangle ~~~ {\rm and} ~~~
\overline{T}^{3,flow}_{abcdef} = \langle \partial_a u_b \partial_c u_d \partial_e u_f \rangle
\label{eq:def_T2T3}
\end{equation}
where the superscript \textit{flow} denotes either TCF, HSF or homogeneous 
isotropic turbulence (HIT) flows. 
The modification induced by a mean shear on
the velocity gradient correlation tensors, defined by Eq.~\eqref{eq:def_T2T3},
are far more intricate than in the case of an axisymmetric
flow, which can be fully analyzed in terms of a small 
number of functions of the radial distance to the axis~\cite{Batchelor46}.
In the TCF, we restrict ourselves to the region which 
is far away from the wall, in the so-called log-layer, where the influence
of the boundary is not too strong~\cite{Lumley,Pope}. 
The number of different 
components of $\overline{T}^{n,flow}$ can be simply estimated from the
9 elements of $\mathbf{A}$ to be equal to $45$ for $n = 2$ and $165$ for 
$n = 3$ - the incompressibility constraint reduces these numbers to
at most $36$ ($92$) independent components for $n = 2$ ($n=3$).
The present work purports to analyse $\overline{T}^{n,flow}$ as a tensor on
its own right, before investigating its particular components. \\ 
An obvious point of comparison
for these tensors is provided by the simpler case of
HIT. For $n = 2$ and $3$, these tensors have a simple
form which can be explicitly written out. An important remark is that
these tensors depend only on 1 dimensional parameter: the mean value of 
$ \Sigma_{ab} \langle \partial_a u_b \partial_a u_b \rangle = 
\langle {\rm tr}( \mathbf{A} \mathbf{A}^T ) \rangle$ for $n = 2$,
and 
$\langle {\rm tr} (\mathbf{S}^3) \rangle$, where $\mathbf{S}$ is the symmetric part
of $\mathbf{A}$: $\mathbf{S} = \frac{1}{2} ( \mathbf{A} + \mathbf{A}^T )$.
We recall that $\langle {\rm tr} ( \mathbf{A} \mathbf{A}^T ) \rangle$ is,
up to viscosity, equal to the dissipation rate of kinetic energy in the 
fluid, 
whereas $\langle {\rm tr} ( \mathbf{\mathbf{S}}^3 ) \rangle$ is up to an 
immaterial numerical factor the rate of production of small scales
(vortex stretching),  
for all the flows considered here.
Dividing the tensors
$\overline{T}^{2,flow}$ by $\langle {\rm tr} ( \mathbf{A} \mathbf{A}^T ) \rangle$
and
$\overline{T}^{3,flow}$ by $\langle {\rm tr} ( \mathbf{S}^3 ) \rangle$
leads to dimensionless forms of the tensors, which can be compared to one 
another.

Parity considerations suggest to decompose the tensors $\overline{T}^{n,flow}$ 
into even and odd components.
Elements of the tensors with at least one odd number of indices equal to 
$1$, $2$ or $3$ should vanish automatically in the presence of an isotropic
forcing, and may only be nonzero because of the anisotropic forcing (the shear).
Decomposition $\overline{T}^{n,flow}$ as the sum of an even and an odd 
contribution is a very natural way to analyse the properties of 
$\overline{T}^{n,flow}$.

Here, we quantify how anisotropic is the flow by comparing the structure of 
the second and third order velocity tensors of the HSF and TCF with the 
corresponding HIT structures. In practice, we simply do a straighforward
least square fit of $\overline{T}^{n,flow}$ of the form 
$\overline{T}^{n,flow} = \overline{\zeta} \times \overline{T}^{n,HIT} + \overline{\Theta^{n,flow}}$. 
The numerical results show that the dimensionless coefficient 
$\overline{\zeta}$ is equal to the ratio of the quantities 
$\langle {\rm tr} ( \mathbf{A} \mathbf{A}^T \rangle $, for $n = 2$, and
$\langle {\rm tr} ( \mathbf{S}^3 )  \rangle $, for $n = 3$, corresponding to
the two flows.
The norm of 
$\overline{\Theta}$ provides a direct measure of the departure from isotropy.
In addition, we can compare the deviation 
$\overline{\Theta}^{n,flow}$ between different flows, in particular between
TCF and HSF, performing again a least square fit analysis. This leads us
to the conclusion that the general structures of 
$\overline{\Theta^{n,flow}}$ are very close to each other for the two shear 
flows considered (HSF, TCF).

This article is organized as follows. The numerical data used in this work is
briefly presented in Section~\ref{sec:num_data}.
The structure of the tensors $\overline{T}^{n,HIT}$, for $n = 2$ and $3$
in the case of HIT flows, as well as the general method we used to process
our shear flow data, are given in Section~\ref{sec:HIT}. 
The lack of isotropy in turbulent shear flows is discussed in terms of
the deviation, $\overline{\Theta}^{n,flow}$, between 
$\overline{T}^{n,flow}$ and $\overline{T}^{n,HIT}$ in
Section~\ref{sec:comp_HIT_shear}. These deviations are compared between
HSF and TCF in Section~\ref{sec:comp_HSF_TCF}.
We then discuss the components of the tensor $\overline{\Theta}^{n,flow}$, 
focusing on the largest ones, see Section~\ref{sec:structure}. Last, we 
recapitulate and discuss our results in Section~\ref{sec:discussion}

\section{DNS data }
\label{sec:num_data}

The analysis presented in this work is based on the channel flow 
simulations~\cite{Graham14}
made publicly available on the Turbulence Database from the Johns Hopkins 
University~\cite{Li2008}, on the one hand, and
on numerical simulations of HSF at two different (moderate) resolutions, 
carried out on the workstations in the Physics Laboratory at the ENS Lyon,
on the other hand.

We use here the standard convention and denote by $x$, $y$ and $z$ the
coordinates in the streamwise, normal to the wall, and spanwise directions, 
respectively.

The data used from the analysis of TCF is the same as 
in~\cite{Pumir+16}. Briefly, the total height of the channel is $2 h$, with
$h = 1$. The streamline extent of the simulated domain is $8 \pi h$, and the 
spanwise extend is $3 \pi h$. The Reynolds number of the flow, 
based on the friction velocity at the wall, $u_\tau$ 
is $\Rey_{\tau} = u_\tau h/\nu = 9.997 \times 10^2$. 
As it is customary, the velocity $u_\tau $ is defined in terms of the 
averaged
shear stress at the wall, $\tau_w$, by $ u_\tau \equiv \sqrt{ 2 \tau_w/\rho}$, 
with $\rho$ the fluid density. The distance to the wall is denoted here
by the dimensionless variable $y^+$, defined by $y^+ = y u_\tau/\nu$. In
the flow studied here, the center of the channel is at $y^+ \approx 1000$.

We recorded data in $20$ planes parallel to the wall. In these planes, we
saved the velocity gradient tensor over a uniform grid of size $\Delta x = 
\Delta z = \pi h /200$, covering the entire simulation domain 
$0 \le x \le 8 \pi h$ and $0 \le z \le 3 \pi h$, corresponding to $9.6 \times
10^5$ points per plane. We collected data at 10 different times, separated
by a time interval of $2$, so the data shown here corresponds to an average
over $\sim 10^7$ data points.
The quality of the statistics has been documented in \cite{Pumir+16}.

In addition, we ran HSF simulations at a resolution $320\times 160 \times 160$
and $200\times 100 \times 100$, corresponding to spatial domains
of size $4 \pi \times 2 \pi \times 2 \pi$, which corresponds to a range of
parameters where HSF simulations are free from box effects, and provide
good models for shear-driven turbulence~\cite{Sekimoto16}.
The code was described in~\cite{PumShr95,Pumir96}. The Reynolds number of the
flows are $\Rey = S (2 \pi)^2/\nu \approx 10^4 $ 
(respectively $\Rey \approx 5.6 \times 10^3$) at the highest 
(respectively lowest) resolution. The flows are adequately
resolved, with a value of the product $k_{max} \eta \approx 1.3-1.4 $.
The velocity gradients were calculated on the collocation points, using
spectral accuracy. The statistics were 
accumulated over very long times: $T_{stat} =  432\times S^{-1}$ (respectively
$720 \times S^{-1}$) at the highest (respectively lowest) Reynolds number, which
corresponds to at least 10 bursts of the kinetic energy, recorded over the 
whole system, according to the mechanisms described 
in~\cite{Pumir96,Sekimoto16}.
We define here the Reynolds numbers based on the 
Taylor microscale by using the velocity fluctuation,
and its derivative along the streamwise direction:
$R_\lambda = \langle u_x^2 \rangle/[ \nu \langle (\partial_x u_x)^2 \rangle^{1/2}]$. 
This corresponds to the Reynolds number routinely measured in laboratory
wind-tunnel experiment~\cite{GargWarh98,ShenWarh00}. The values
found here are $R_\lambda \approx 145$ and $120$ for the two flows.
The value of $R_\lambda$ is comparable to the value found in the TCF, 
in the range $200 \lesssim y^+ \lesssim 600$, see Fig.1b of \cite{Pumir+16}.

In both HSF and TCF, the velocity field is decomposed as a mean 
flow, $U(y) \mathbf{e}_x$, plus a fluctuation term, $\mathbf{u}$. 
Throughout this text, $\mathbf{u}$ refers to the 
fluctuation of the velocity field. 
In the TCF case, the properties of the turbulent
velocity fluctuations depend on the distance to the wall.

For the sake of completeness, we also investigated the velocity gradient tensor
in HIT. To this end, we used the data at $R_\lambda \approx 275$, discussed 
in ~\cite{PXBG16}. The data was accumulated over $\approx 2$ eddy turnover
times.

To check the reliability of the results presented here, 
the analysis presented below was done both with the entire dataset, 
and repeated with only one half of it, corresponding to the first half 
of the runs. The results
presented below, obtained with the entire dataset, are found to differ only 
slightly from those obtained with only half the dataset.
In the following, we indicate how the data obtained with the full dataset
compares with that obtained with only half the dataset. Overall, 
these comparisons
give us confidence that the results presented here are very reliable.

\section{Method of analysis }
\label{sec:HIT}

In the rest of the text, $X_1$, $X_2$ and $X_3$ refer to the components
of the vector $\mathbf{X}$ in the $x$, $y $ and
$z$ directions, respectively. With this convention, 
the HSF problem, the fluctuation around 
the mean shear is $\partial_2 u_1$.
The second and third order correlations of
the velocity gradient tensor in various shear flows are defined by 
Eq.~\eqref{eq:def_T2T3}.

 A useful starting point 
to study these correlation functions is provided by the simpler HIT case.
The comparison with HIT flows is particularly relevant, since
turbulent flows
at extremely 
high Reynolds numbers are expected to recover isotropic properties.
In this section, we briefly recall the expressions for the second and
results of Direct Numerical Simulations (DNS) for HIT flows. 
We also explain how we systematically compare various flows.

\subsection{Second order tensor} 

In a homogeneous and isotropic flow, the expression of the second order 
tensor function, $\langle \partial_a u_b \partial_c u_d \rangle$,
can be simply obtained from elementary considerations~\cite{Landau}. Namely,
the tensor can be expressed only in terms of the Kronecker $\delta$-tensor,
and symmetry imposes that:
\begin{equation}
\overline{T}^{2,HIT}_{abcd} \equiv 
\langle \partial_a u_b \partial_c u_d \rangle  = A \delta_{ac} \delta_{cd} + 
B \delta_{ad} \delta_{bc} + C \delta_{ab} \delta_{cd} 
\label{eq:general_2d}
\end{equation}
We use the Einstein convention of summation of repeated indices throughout.
Incompressibility imposes that
$\langle \partial_a u_a \partial_b u_c \rangle = 0$, so 
$A + B + 3C = 0$. In addition, homogeneity imposes that 
$\langle \partial_a u_b \partial_b u_a \rangle = 0$~\cite{Betchov56}, 
which leads to:
$A + 3B + C = 0$. Last, the dissipation of kinetic energy is equal
to $\nu \langle \partial_a u_b \partial_a u_b \rangle = \varepsilon$,
which gives rise to: $3A + B + C = \frac{\varepsilon}{3 \nu}$.
This leads to the explicit expression for 
$\langle \partial_a u_b \partial_c u_d \rangle$:
\begin{equation}
\overline{T}^{2,HIT}_{abcd} = \frac{\varepsilon}{30 \nu} ~ \times ~  T^{2,HIT}_{abcd} ~~ {\rm with: } ~~~
T^{2,HIT}_{abcd} \equiv  \Bigl( 4 \delta_{ac} \delta_{bd} - \delta_{ab} \delta_{cd} - \delta_{ad} \delta_{bc} \Bigr)
\label{eq:dvdv_hit}
\end{equation}
The second order moment velocity gradient tensor is therefore expressed in 
terms of only one dimensional quantity, 
$\varepsilon/\nu = \langle \partial_a u_b \partial_a u_b \rangle$.

We note that only the elements of the tensor $T^{2,HIT}_{abcd}$ which contain
an even number of $1$, $2$ and $3$ among the indices are nonzero. 

\subsection{Third order tensor} 

The third order velocity gradient tensor, $\overline{T}^{3,HIT}$,
can be expressed using similar principles~\cite{Pope}. 
The calculation, presented
in the Appendix, leads to the relation:
\begin{equation}
\overline{T}_{abcdef}^{3,HIT} = \langle {\rm tr}(\mathbf{S}^3) \rangle ~ \times ~ T^{3,HIT}_{abcdef} \label{eq:intro_T3}
\end{equation}
where $\mathbf{S}$ is the rate of strain tensor: 
$S_{ab} = (\partial_a u_b + \partial_b u_a)/2$. 
As it was the case
for the second order moment velocity gradient tensor, and as
observed by~\cite{Siggia81}, the third order moment velocity gradient tensor 
for a homogeneous, isotropic turbulent flow is expressible in terms of only
one dimesional quantity, namely $\langle {\rm tr}(\mathbf{S}^3) \rangle$. 
Namely, the tensor $T^{3,HIT}$ reads:
\begin{eqnarray}
T^{3,HIT}_{abcdef} 
&=& \bigg\{ \frac{8}{105} \delta_{ab} \delta_{cd} \delta_{ef} \nonumber \\
&- & \frac{2}{35} [ 
\delta_{ab} ( \delta_{ce} \delta_{df} + \delta_{cf} \delta_{de} ) +
\delta_{cd} ( \delta_{ae} \delta_{bf} + \delta_{af} \delta_{be} ) +
\delta_{ef} ( \delta_{ac} \delta_{bd} + \delta_{ad} \delta_{bc} )  ] \nonumber \\
& + & \frac{3}{70} (
\delta_{ac} \delta_{de} \delta_{fb} + \delta_{ac} \delta_{df} \delta_{eb} +
\delta_{ad} \delta_{ce} \delta_{fb} + \delta_{ad} \delta_{cf} \delta_{eb} \nonumber \\
& & ~~ +\delta_{bc} \delta_{de} \delta_{fa} + \delta_{bc} \delta_{df} \delta_{ea} +
\delta_{bd} \delta_{ce} \delta_{fa} + \delta_{bd} \delta_{cf} \delta_{ea} ) \nonumber \\
& + & \frac{1}{45} \delta_{ef} ( \delta_{ac} \delta_{bd} - \delta_{ad} \delta_{bc} ) + \frac{1}{45}
\delta_{cd} ( \delta_{ae} \delta_{bf} - \delta_{af} \delta_{be} ) + \frac{1}{45}
\delta_{ab} ( \delta_{ce} \delta_{df} - \delta_{cf} \delta_{de} ) \nonumber \\
& - & \frac{1}{30} (\epsilon_{abe} \epsilon_{cdf} + \epsilon_{abf} \epsilon_{cde} ) 
- \frac{1}{30} ( \epsilon_{abc} \epsilon_{efd} + \epsilon_{abd} \epsilon_{efc} ) 
- \frac{1}{30} ( \epsilon_{cda} \epsilon_{efb} + \epsilon_{cdb} \epsilon_{efa} ) \bigg\} 
\label{eq:dvdvdv_hit}
\end{eqnarray}
The expression ~\eqref{eq:dvdvdv_hit} shows that 
the components of $T^{3,HIT}_{abcdef}$ are automatically zero if 
the number of any single index among $(a,b,c,d,e,f)$ is odd.

\subsection{Comparison of turbulent shear flows with HIT flows}
\label{subsec:comp_flows}

 To estimate how close to isotropy is the small-scale structure of the flow,
we compare systematically the second
and third moments of the turbulent velocity gradient tensor with 
their form, in the case of a homogeneous and isotropic flow. 
These tensors, Eq.~\eqref{eq:dvdv_hit} and Eq.~\eqref{eq:dvdvdv_hit}, depend
on only one dimensional parameter, with an otherwise completely determined
structure. 

To take advantage of this structure, for all flows considered, we determine
the invariants $C_2 = \langle \partial_a u_b \partial_a u_b \rangle$, and 
$C_3 =  \langle {\rm tr}(\mathbf{S}^3 ) \rangle$, and divide the expressions 
of the second
and third order tensors, determined numerically, by $C_2$ and $C_3$, 
respectively. In the TCF case, the dependence of the turbulence 
properties on the distance to the wall, implies that both
$C_2$ and $C_3$ depend on $y^+$.
The normalized expressions of these tensors, 
$T^{2,flow}$ and $T^{3,flow}$, are defined by:
\begin{equation}
T^{2,flow} = \frac{\overline{T}^{2,flow}}{\langle \partial_a u_b \partial_a u_b \rangle } ~~~ {\rm and } ~~~
T^{3,flow} = \frac{\overline{T}^{3,flow}}{\langle tr(s^3) \rangle } 
\end{equation}
With these definitions, $T^{2,flow} $ satisfies $T^{2,flow}_{abab} = 1$,
whereas $T^{3,flow}$ is constrained by
a condition expressing that $\langle {\rm tr}(\mathbf{S}^3) \rangle = 1$.
The dimensionless expressions can be directly compared
with those of $T^{2,HIT}$ and $T^{3,HIT}$,
Eq.~\eqref{eq:dvdv_hit} and \eqref{eq:dvdvdv_hit}, and with one another.

 With a simple least square minimizing technique, we find
the minimum of $|| T^{n,flow} - \zeta \times T^{n,HIT}||^2$, where
$|| . ||^2$ is the usual Euclidean norm: 
$|| T_{abcd} ||^2 = \sum_{abcd} T_{abcd}^2 $. This allows us to estimate how 
close the tensor $T^{n,flow}$ is to $T^{n,HIT}$: the determined
value of $\zeta$ is such that 
$T^{n,flow} = \zeta \times T^{n,HIT} + \Theta^{n,flow}$, where $T^{n,HIT}$
and $\Theta^{n,flow}$ are orthogonal to each other, and 
$|| \Theta^{n,flow}||^2$ provides a quantitative measure of how much
$T^{n,flow}$ differs from isotropy.

The symmetry of the problem suggests to decompose the 
tensors $T^{n,flow}$ and $\Theta^{n,flow}$ ($n = 2$, $3$), as a sum of 
an even and an odd part. 
The even component, denoted $\Theta_{ev}^{n,flow}$, corresponds to a tensor
with an even number of $1$, $2$ and $3$ among all the indices.
In a flow which is invariant under all the symmetries $x_i \rightarrow -x_i$,
the tensor $\Theta^{n,flow}$ is necessarily even. Such a symmetry is implied
by isotropy.  \\
The odd part, $\Theta_{od}^{n,flow}$ contains all the other elements of 
the tensor, with 
at least one odd number of $1$, $2$ or $3$ among all the indices.
This allows us to write the decomposition:
\begin{equation}
T^{n,flow} = \zeta \times T^{n,HIT} + \Theta_{ev}^{n,flow} + \Theta_{odd}^{n,flow}
\label{eq:decomp_T}
\end{equation}

In the presence of a shear, some components of $\Theta_{odd}^{n,flow}$ may
be nonzero, as we now explain.
Specifically, we write the velocity as
$U(x_2)  \mathbf{e}_1 + \mathbf{u}(\mathbf{x},t)$, where 
$\mathbf{U}(x_2)$ is
the mean flow (we recall that $1$ is the streamwise direction, and 
$2$ the direction normal to the wall in the case of a turbulent channel flow), 
and $\mathbf{u}(\mathbf{x},t)$ is the fluctuation, satisfying
$\langle \mathbf{u}(\mathbf{x}) \rangle = \mathbf{0}$, where the average here is taken
as a time average at any given point in the flow domain. 
It is straightforward to check that the Navier-Stokes equations,
written for the fluctuation $\mathbf{u}$, are invariant under 
reflection in the spanwise direction, characterized by both 
$x_3 \rightarrow - x_3$ and $u_3 \rightarrow - u_3$.
This allows us to restrict ourselves to solutions which are even in the 
spanwise direction, $x_3$: in our problem, individual components of the 
velocity tensor with an odd number of $1$ and of $2$ 
indices may be nonzero, provided the number of $3$ is even.

We note that the method used here to compare the second and third
order correlation tensors with homogeneous, isotropic flows can be readily 
generalized to compare different shear flows among themselves. The corresponding
notation will be specified later, see Section~\ref{sec:comp_HSF_TCF}.

\section{Comparison between homogeneous isotropic flows and shear flows}
\label{sec:comp_HIT_shear}

In this section, we systematically compare the second and third order
velocity gradient correlation tensors, following the 
decomposition \eqref{eq:decomp_T}. Before we proceed to the detailed
analysis of several flows, we notice that the coefficient $\zeta $
in Eq.~\eqref{eq:decomp_T} was always found to be extremely close to $1$,
so we simply denote the tensors of $\Theta_{od}^{n,flow}$ 
and $\Theta_{ev}^{n,flow}$ as the deviations between the correlation tensors
and their homogeneous, isotropic predictions. 

\subsection{Numerical results with a homogeneous isotropic simulation}
\label{subsec:comp_HIT_DNS}

In this subsection, we present results obtained for an explicitly 
homogeneous isotropic flow at moderate Reynolds number, as detailed in
Section~\ref{sec:num_data}, with the aim of testing the analysis applied 
in this work.  \\
In the decomposition given by Eq.\eqref{eq:decomp_T}, we find that the values 
of $\zeta $ differ from 
$1$ by at most $\approx 0.2 \%$, both for the second and third order
velocity gradient correlation tensors.

In all cases, the deviations between the predicted correlation form 
Eq.~\eqref{eq:dvdv_hit} and \eqref{eq:dvdvdv_hit} are very small. More
precisely,
for the second order correlation tensor, the norms of the discrepances 
are $|| \Theta_{ev}^{2,HIT num}||^2/|| T^{2,HIT} ||^2 \approx 2 \times  10^{-4}$ 
and $|| \Theta_{od}^{2,HIT num}||^2/|| T^{2,HIT} ||^2 \approx 4 \times 10^{-5}$.
The errors are slightly larger for discrepancies characterizing 
the third order correlation tensor:
$|| \Theta_{ev}^{3,HIT num}||^2/|| T^{3,HIT} ||^2 \approx 1.2 \times 10^{-3}$ 
and $|| \Theta_{od}^{3,HIT num}||^2/|| T^{3,HIT} ||^2 \approx 1.7 \times 10^{-2}$.
We have monitored here all the possible components of the tensor 
$\Theta_{od}^{n,HIT num}$. 
The relatively large value of the $|| \Theta_{od}^{3,HIT}||^2/|| T^{3,HIT}||^2$ 
can be attributed in parts to terms of the form 
$\langle (\partial_a v_b)^3 \rangle$ (with $a \ne b$), which are found to be 
surprisingly high.
Quantitatively, the sum of the $6$ corresponding terms was found to account to 
more than $20\%$ of the total norm of $|| \Theta_{od}^{3,HIT num}||^2$. \\
The relatively large values of the third moment of $\partial_a u_b$
($a \ne b$), in a flow 
which is expected to be statistically isotropic, are surprising. They 
could conceivably be induced by a local shear
at large scale, persistent for a time of the order of the 
eddy turnover time. As already 
noticed~\cite{PumShr95}, these large scale gradients are sufficient to 
generate large third moments of $\partial_a u_b$ (see also \cite{HolzSig94}). 
We note that these moments
are appreciable in our simulation, which is run only for approximately
$2$ eddy-turnover times. 
It is expected that these moments would eventually average out to zero 
in a simulation run for a much longer time~\cite{Vreman:14a}.

\subsection{Numerical results with homogeneous shear flows }
\label{subsec:comp_HIT_HSF}

The results of the comparison between HSF and HIT are summarized
in Table~\ref{table:HSF_glob}.
The corresponding values of 
$\zeta$ (not shown) are all extremely close to $1$.

\begin{table}
\begin{center}
\begin{tabular}{| c | c | c || c | c |}
\hline
& $||\Theta_{ev}^{2,HSF}||/||T^{2,HIT}||$ & $||\Theta_{od}^{2,HSF}||/||T^{2,HIT}||$ & $\Theta_{ev}^{3,HSF}||/||T^{3,HIT}||$ & $||\Theta_{od}^{3,HSF}||/||T^{3,HIT}||$ \\ \hline \hline
$R_\lambda=120$ & $ 0.12 $ & $0.18$ & $0.23$ & $0.67$  \\ \hline 
$R_\lambda=145$ & $ 0.083 $ & $0.15$ & $0.18$ & $0.58$  \\ \hline 
\end{tabular}
\end{center}
\caption{Norms of the deviations of $T^{2,HSF}$ and $T^{3,HSF}$, as defined
by Eq.~\eqref{eq:decomp_T} and normalized by $|| T^{2,HIT} ||$ and 
$|| T^{3,HIT}||$, respectively. The values shown here were obtained
by using the full dataset; they differ from the values
obtained with half the dataset (as explained in Section~\ref{sec:num_data}),
by no more than $5 \%$. 
}
\label{table:HSF_glob}
\end{table}

The deviations of the second order correlation tensor from the 
homogeneous isotropic predictions are relatively weak. The odd contribution,
$|| \Theta_{od}^{2,HSF} || $, is larger than the even one,
$|| \Theta_{ev}^{2,HSF} || $ by roughly $ 50 \%$.
Both the odd and even components decay
slightly when the Reynolds number increases, possibly like $R_\lambda^{-1}$,
which is the prediction based on elementary 
arguments~\cite{Corrsin58,Lumley67}, and consistent with previous 
numerical work~\cite{Pumir96}.  \\
In comparison, the deviations
measured for the third order velocity derivative correlation tensor
are much larger. The odd component is significantly larger
than the even one, roughly by a factor of $ \gtrsim 3$, irrespective of the 
Reynolds number. 
The decay of $|| \Theta_{od}^{3,HSF} ||/|| T^{3,HIT} ||$ with the 
Reynolds number is certainly slower than $R_{\lambda}^{-1}$, 
and is possibly consistent with 
the power law $R_\lambda^{-1/2}$ found in~\cite{ShenWarh00}. 

\subsection{Numerical results with turbulent channel flows }
\label{subsec:comp_HIT_TCF}

\begin{figure*}
\begin{center}
\subfigure[]{
        \includegraphics[width=0.45\textwidth]{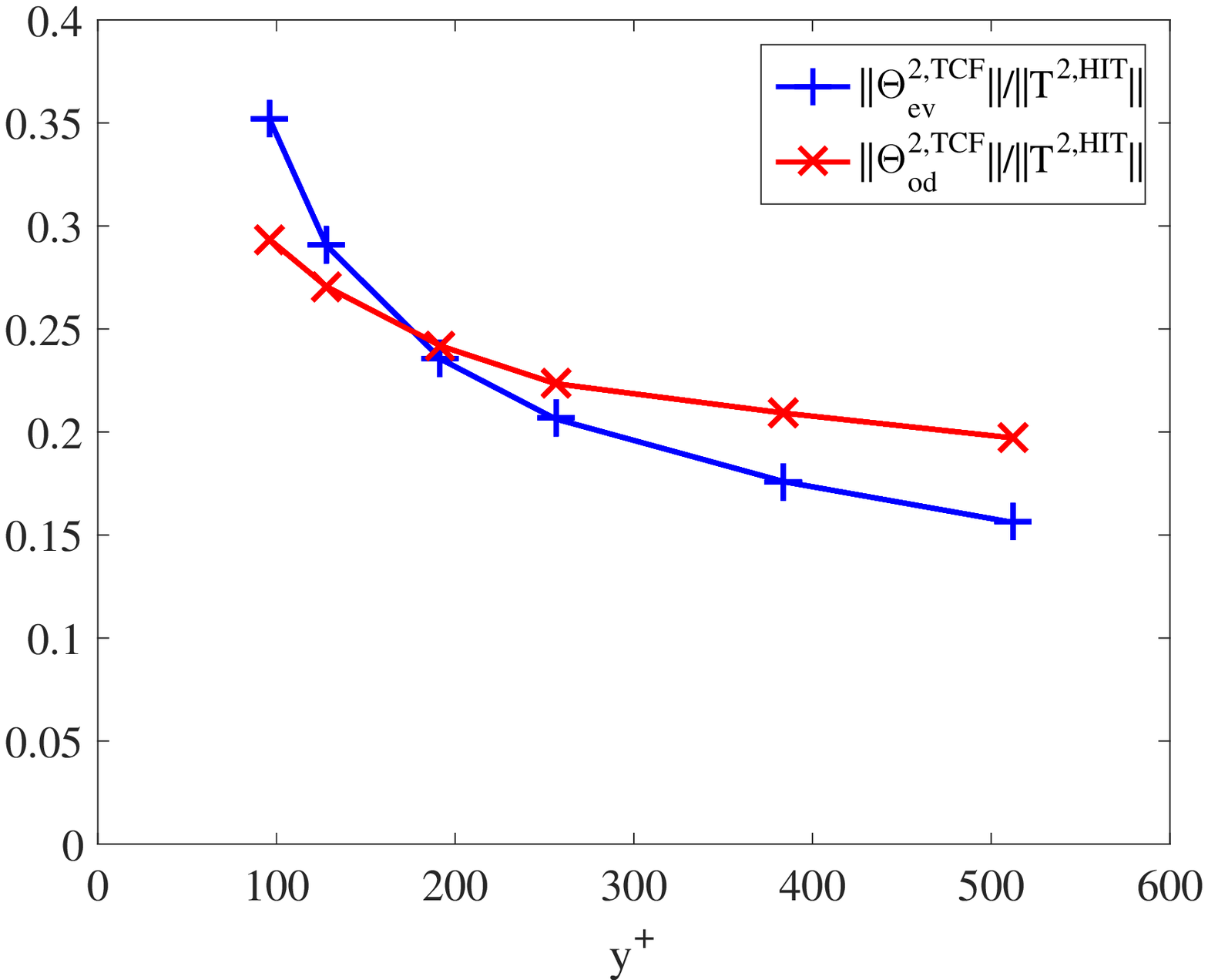}
}
\subfigure[]{
        \includegraphics[width=0.45\textwidth]{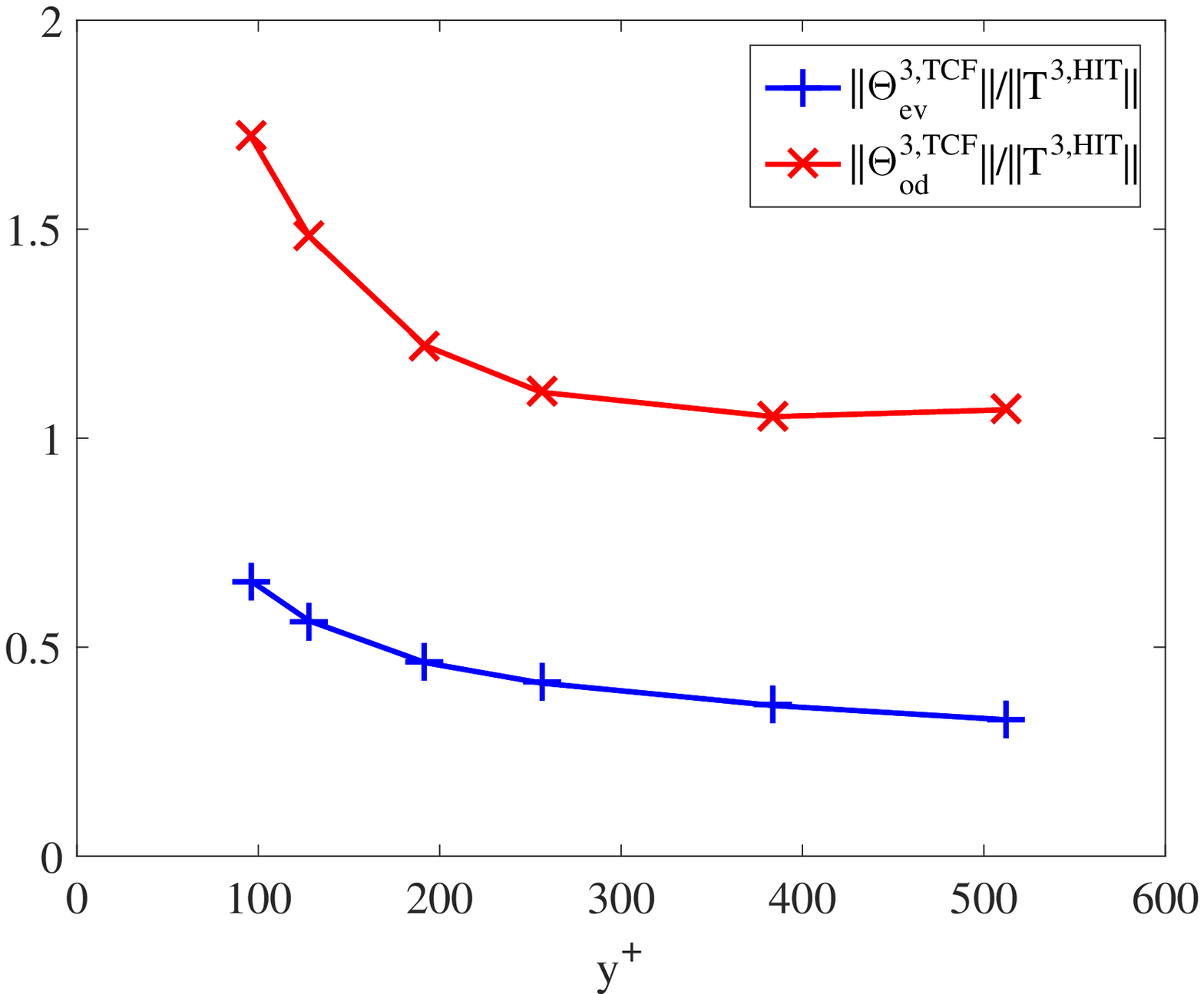}
}
\caption{The norm of the deviation from the HIT tensor, 
$|| \Theta_{ev}^{n,TCF} ||/|| T^{n,HIT}||$ ("+" symbols)
and
$|| \Theta_{odd}^{n,TCF} ||/|| T^{n,HIT}||$ ("x" symbols)
for $n = 2 $ (left panel)
and $n = 3$ (right panel), as a function of the distance from the wall $y^+$.
The center of the channel is at $y^+ \approx 1000$.
The values shown here were obtained by processing the whole dataset.
Using only half the dataset, as explained in Section~\ref{sec:num_data},
would result in differences by no more than $5 \%$. 
}
\label{fig:comp_TCF_HIT}
\end{center}
\end{figure*}

Over the entire channel flow, the values of $\zeta$ in Eq.~\ref{eq:decomp_T} 
differ from $1$ by no more than a few percent, even very close to the wall.
The results for the norms of the deviations between the second and third order
velocity gradient correlation tensors, and the isotropic predictions are
presented in Fig.~\ref{fig:comp_TCF_HIT}, which shows
the values of $||\Theta_{ev}^{n,TCF}||$ and
$||\Theta_{od}^{n,TCF}||$ ($n=1$, $2$), normalized by $|| T^{n,HIT} ||$,
and measured at several values of $y^+$
in a region, $100 \lesssim y^+ \lesssim 500$ (the center of the channel
is at $y^+ \approx 1,000$). 
The range of values of $y^+$ shown
corresponds essentially
to the so-called logarithmic layer of the turbulent channel flow.
In this region, 
the second (left panel) and third order (right panel) velocity gradient 
tensors deviate significantly
more from the homogeneous and isotropic predictions, 
than the HSF flow, compare with Table~\ref{table:HSF_glob}. \\
Quantitatively, the deviations $\Theta^{2,TCF}$, normalized by 
$|| T^{2,HIT}||$, shown in the left panel 
of Fig.~\ref{fig:comp_TCF_HIT}, are of the order of $\approx 0.2$, and
decay slightly when $y^+$ increases. The odd component 
$|| \Theta_{od}^{2,TCF}||$ ("x" symbols) is larger than the even component, 
$||\Theta_{ev}^{2,TCF}||$ ("+" symbols) only for $y^+ \gtrsim 200$. 
The norms of $\Theta^{3,TCF}$
divided by $|| T^{3,HIT}||$
are shown in the right panel of Fig.~\ref{fig:comp_TCF_HIT}. 
The norm of the odd part of the tensor $\Theta^{n,TCF}$ values of 
$|| \Theta_{ov}^{n,TCF} ||/|| T^{n,HIT}|| $ very close to the wall are 
much larger than those shown in Fig.~\ref{fig:comp_TCF_HIT}, a simple 
consequence of the very 
strong influence of the boundary for $y^+ \lesssim 100$. 

Numerically, the norm of $\Theta_{od}^{3,TCF}$ normalized
by $|| T^{3,HIT}||$, $||\Theta_{od}^{3,TCF}||/ || T^{3,HIT}||$,
goes up to values $\approx 45 $ at $y^+ = 1$.
For $y^+ \gtrsim 200$, in comparison, as shown in the right panel
of Fig.~\ref{fig:comp_TCF_HIT} ("x" symbols), the values of 
$|| \Theta_{od}^{3,TCF}||/||T^{3,HIT}||$
are approximately constant, and equal to $\approx 1.1$, a value roughly
$\lesssim 2$ times larger than in HSF at $R_\lambda = 145$. The values 
of $|| \Theta_{ev}^{3,TCF}||/||T^{3,HIT}||$ ("+" symbols in the right
panel of Fig.~\ref{fig:comp_TCF_HIT}) vary in relative value similarly
to those of $|| \Theta_{od}^{3,TCF}||/||T^{3,HIT}||$.
Close to the wall, 
at $y^+ = 1$, the measured value is $\approx 3$. In the log-layer,
the values are $\approx 0.3-0.4$. We notice that the values shown
in Fig.~\ref{fig:comp_TCF_HIT} are also roughly $\lesssim 2$ times larger than 
those found in the case of the HSF. 

Fig.~\ref{fig:comp_TCF_HIT} and the discussion so far
have been focused on the structure of the flow 
mostly in the log-layer of the channel,
where the shear is significant.
We end this subsection by noticing that, at the center of the 
channel, at $y^+ \approx 1000$ (data not shown in Fig.~\ref{fig:comp_TCF_HIT}),
the odd components of the
deviations $\Theta_{od}^{n,TCF}$ vanish. In fact, the value of 
$||\Theta_{od}^{n,TCF}||^2$ are comparable to (in fact, even
smaller than) the values found when comparing a HIT DNS with the 
theoretical prediction, discussed in Subsection \ref{subsec:comp_HIT_DNS}. In
comparison, the values of $|| \Theta_{ev}^{n,TCF}||^2 $ are found to be 
significantly larger than those listed in Subsection \ref{subsec:comp_HIT_DNS}:
$|| \Theta_{ev}^{2,TCF}||^2  \approx 10^{-2}$ and 
$|| \Theta_{ev}^{3,TCF}||^2  \approx 8. \ 10^{-2}$.

\section{Comparison between Homogeneous shear flows and turbulent channel flow}
\label{sec:comp_HSF_TCF}

Having explored in the previous section the magnitude of the deviations
$\Theta^{2,3}$ in the case of HSF and TCF, we now ask 
how similar are the deviations of the second and third
order velocity gradient tensor from their predicted form in the case of
a HIT flow, Eq.~\eqref{eq:dvdv_hit} and \eqref{eq:dvdvdv_hit}. Specifically,
to compare two flows, flow 1 and flow 2,
we represent the tensor corresponding to flow 1, 
$\Theta_{ev,od}^{n,flow1}$ ($n = 1$, $2$) 
as a form
proportional to $\Theta_{ev,od}^{n,flow2}$, plus a discrepancy, $\Psi_{ev,od}^{n}$:
\begin{equation}
\Theta_{ev,od}^{n,flow 1} = \mu_{ev,od}^n \Theta_{ev,od}^{n,flow 2} + \Psi_{ev,od}^n
\label{eq:flow_1_2_fit}
\end{equation}
and choose $\mu_{ev,od}^n$ to minimize the norm of the error, 
$|| \Psi_{ev,od}^n||$.
The norm of $\Psi_{ev,od}^n$, normalized with 
$|| \Theta_{ev,od}^{n,TCF}||$, provides a quantitative measure of how close
are the deviations from isotropy in the two flows considered proportional 
to one another.

\subsection{ Comparison between two HSF at different Reynolds numbers}
\label{subsec:comp_HSF_HSF}

The comparison between the two turbulent HSF at $R_\lambda = 145$ and 
$R_\lambda = 120$, shows that the structures of the correlation tensors are
in fact extremely similar. 
In the present subsection, 
we distinguish the deviations $\Theta_{ev,od}^{n,HSF}$ between the two flows
by specifying the Reynolds numbers in the superscript:
$\Theta_{ev,od}^{n,HSF;R120}$ ($\Theta_{ev,od}^{n,HSF;R145}$)
corresponds to $R_\lambda = 120 $ ($R_\lambda = 145$). The present analysis
is based on the decomposition:
\begin{equation}
\Theta_{ev,od}^{n,HSF;R145 } = \mu_{ev,od}^n \Theta_{ev,od}^{n,HSF;R120} + 
\Psi_{ev,od}
\label{eq:comp_HSF_HSF}
\end{equation}
\\ 
The values of the coefficients 
$\mu_{ev,od}^{n}$ and of the ratios 
$|| \Psi_{ev,od}^{n}||/||\Theta_{ev,od}^{n,HSF;R145}||$, 
as defined by Eq.~\eqref{eq:comp_HSF_HSF}, are
shown in Table~\ref{table:comp_HSF_HSF}.
Numerically, the measured values of $\mu^{n}_{ev,od}$, are 
all less than $1$, in the range $0.7-0.8$. 
The small ratio between the norm of
the discrepancy from the linear relation, $|| \Psi_{ev,od}^n||$, and
$||\Theta_{ev,od}^{n,HSF;R145}||$, see Table~\ref{table:comp_HSF_HSF},
suggest that the tensors are close to being proportional to each other. 
This justifies the observation that the values of $\mu_{ev,od}^n$
satisfy:
$\mu^{n}_{ev,od} \approx || \Theta_{ev,od}^{n,HSF;R145}||/
|| \Theta_{ev,od}^{n,HSF;R120}||$ (values in Table~\ref{table:HSF_glob}). 
We notice that the values of $\mu$ are larger for $n = 3$ than for $n= 2$,
which is consistent with the observation of a slower decay of the deviations
from isotropy of the third order correlation functions already 
noticed~\cite{Pumir96}.

In the rest of the text, we will only consider the HSF at 
$R_\lambda = 145$, and simply denote $\Theta_{ev,od}^{n,HSF}$ 
the deviation from the isotropic form of the tensor.

\begin{table}
\begin{center}
\begin{tabular}{ | c | c | c || c | c |}
\hline
& $\mu_{ev}^{n}$ & $||\Psi_{ev}^{n,HSF}||/||\Theta_{ev}^{n,HSF;R145}||$ & $\mu_{od}^{n,HSF}$ & $||\Psi_{od}^{n,HSF}||/||\Theta_{od}^{n,HSF;R145}||$ \\ \hline \hline
$n=2$ & $ 0.72 $ & $0.078$ & $0.80$ & $0.051$  \\ \hline 
$n=3$ & $ 0.77 $ & $0.14$ & $0.86$ & $0.080$  \\ \hline 
\end{tabular}
\end{center}
\caption{Comparison between the two HSF at different Reynolds numbers. 
The notation refers to the Eq.~\ref{eq:comp_HSF_HSF} in the main text.
The data shown here was obtained by processing the full dataset.
The values of $\mu_{ev,od}^{n}$ obtained by processing only half the dataset
(see Section \ref{sec:num_data}) differ by no more than $3\%$, whereas
the small values of the norms of the discrepancy 
$|| \Psi_{ev,od}^{n,HSF} || $ differ by $\approx 15 \%$.
}
\label{table:comp_HSF_HSF}
\end{table}

\subsection{ Comparison between TCF and HSF }
\label{subsec:comp_TCF_HSF}

\begin{figure*}
\begin{center}
\subfigure[]{
        \includegraphics[width=0.45\textwidth]{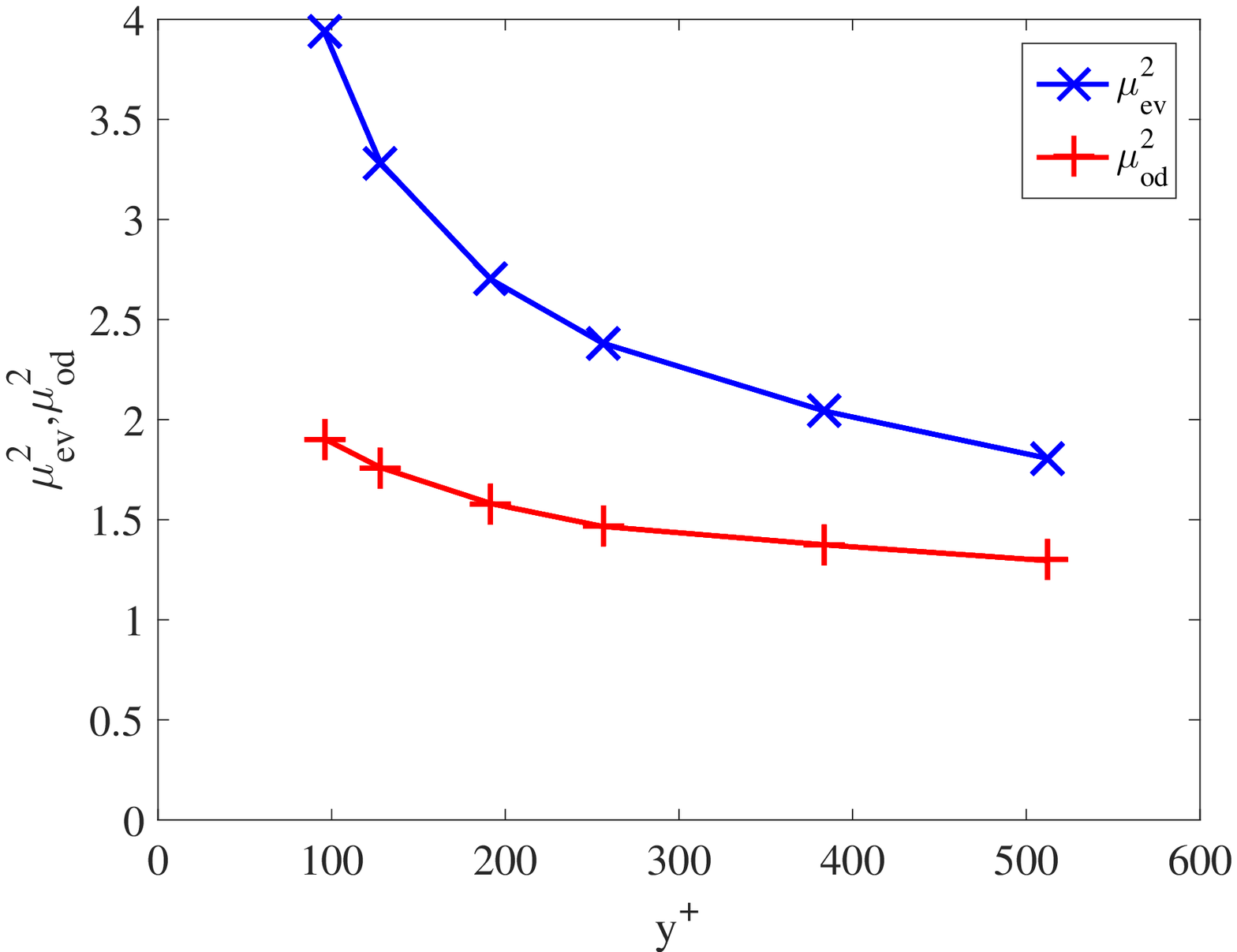}
}
\subfigure[]{
        \includegraphics[width=0.45\textwidth]{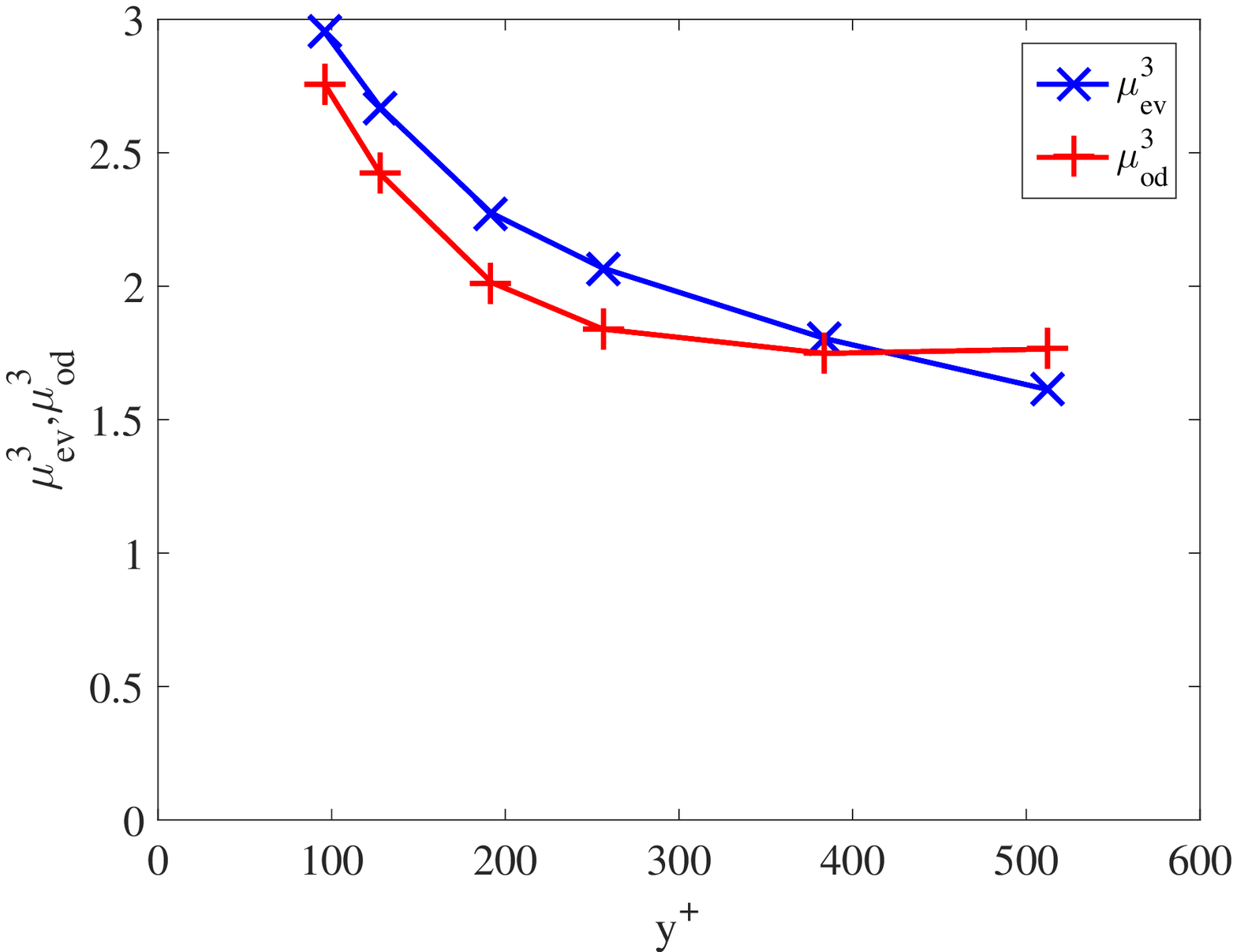}
}
\subfigure[]{
        \includegraphics[width=0.45\textwidth]{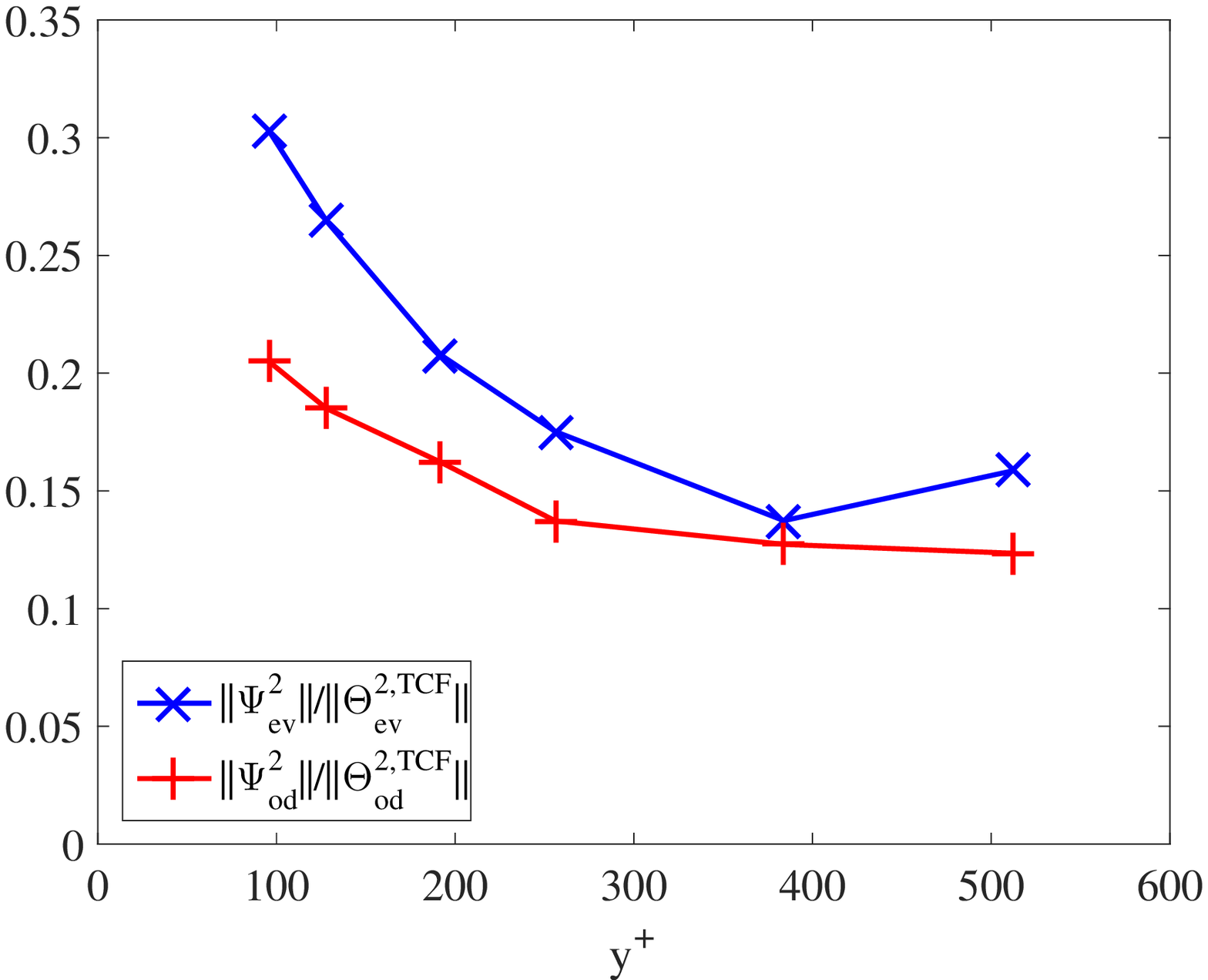}
}
\subfigure[]{
        \includegraphics[width=0.45\textwidth]{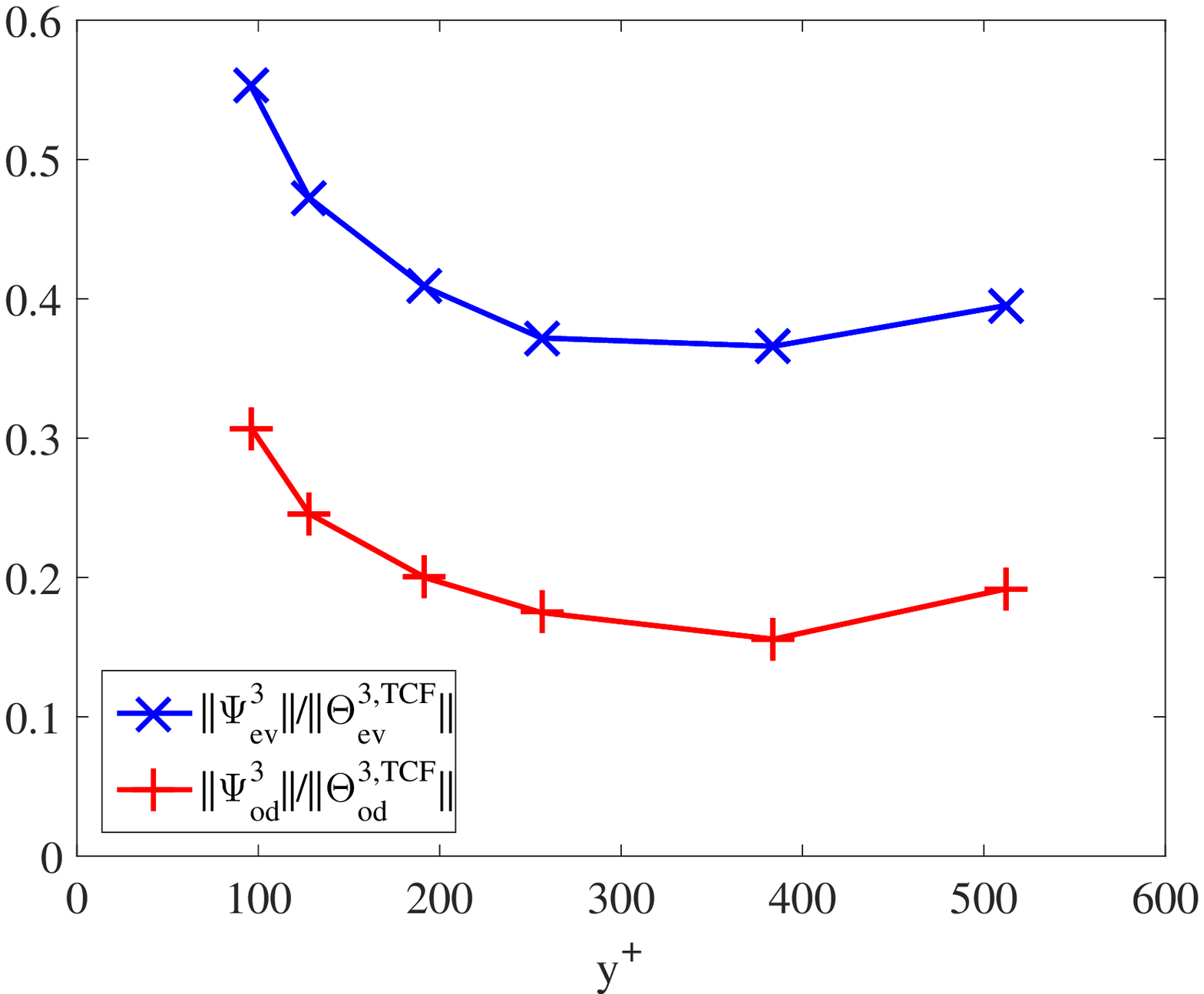}
}
\caption{Comparison between $\Theta_{ev,od}^{n,TCF}$ in the log-layer
of the TCF with $\Theta_{ev,od}^{n,HSF}$ ($R_\lambda = 145$), for
$n = 2$ (left column) and $n = 3$ (right column). The coefficient 
$\mu$, defined in Eq.~\eqref{eq:flow_1_2_fit}, is shown in the upper row,
and the norm of the discrepancy, $\Psi_{ev,od}^{n,HSF}$, divided by
$|| \Theta_{ev,od}^{n,TCF} ||$ in the lower row.  The errors are comparatively 
larger for $n = 3$, especially for the even terms.
The curves shown here were obtained by processing the full dataset.
The values of $\mu$, obtained with only half the 
dataset (see Section ~\ref{sec:num_data}) differ by no more than $5\%$ from
those shown in the figure. The discrepancies 
$|| \Psi_{ev,od}^{n,HSF}||$ also differ by less than $ \approx 5 \%$,
except at $y^+ = 384$ and $y^+ = 512$, where $|| \Psi_{od}^{3,HSF}||$ was
found to differ by $15 \%$ and $10\%$, respectively, from the indicated
values.
}
\label{fig:comp_HSF_TCF}
\end{center}
\end{figure*}

The comparison between the HSF at $R_\lambda = 145$ and the TCF, in the
log-layer, are shown in Fig.~\ref{fig:comp_HSF_TCF}, for the second (left part)
and third (right part) correlation function of the velocity gradient tensor. \\
The values of $\mu$, upper part of Fig.~\ref{fig:comp_HSF_TCF}, generally
vary little in the range $200 \lesssim y^+ \lesssim 500$. This is consistent
with the notion that in the log-layer, the properties of turbulence acquire
some universal character. The values of $\mu^{n}_{ev,od}$ are all found to be
$\approx 2$, pointing to a significantly larger deviation from isotropy in
the case of TCF than of HSF~\cite{Pumir+16}. \\
The relative discrepancies, $||\Psi_{ev,od}^n||/||\Theta_{ev,od}^{n,TCF}||$,
shown in the bottom row of Fig.~\ref{fig:comp_HSF_TCF}, 
are generally low, $\approx 0.15 $, except for the even component 
associated with the third order correlation function of the velocity gradient,
which is as high as $\approx 0.40 $. 

\section{Structure of the third order correlation tensor}
\label{sec:structure}

Table~\ref{table:HSF_glob} and Fig.~\ref{fig:comp_TCF_HIT} show that the
largest deviations from isotropy are those measured by the third order
velocity gradient correlation tensor, $T^{3,HSF}$ and $T^{3,TCF}$, the 
deviations from the HIT predictions being approximately twice smaller for the
latter than for the former in the log-layer of the boundary layer.
In this section, we focus on the largest components of 
$\Theta_{od,ev}^{3,TCF}$ and $\Theta_{od,ev}^{3,HSF}$.  

Tables~\ref{table:comp_od_HSF_TCF} and 
\ref{table:comp_ev_HSF_TCF}, discussed in more detail in the following
subsections, list some of the components of the tensors
$\Theta_{od,ev}^{3,HSF} $ and  
$\Theta_{od,ev}^{3,TCF} $ at $y^+ = 256 $ and $y^+ = 512$, including
the largest components measured numerically. 
As explained in Section~\ref{sec:num_data}, we systematically compared
the data shown below, obtained with the full dataset, with the values obtained
with only half the dataset.
In almost all cases where the absolute value of the listed component is 
larger than $0.1$, the results were found to differ in the second significant
figure by no more than $\pm 1$, very rarely by $\pm 2$. The 
figure indicated for very small coefficients
(less than $0.1$) were found in most cases to
be unchanged when using half the dataset, or to differ by at most $1$, and by 
$2$ in only a very few cases. None of the signs of the quantities listed in 
the table was found to change when processing only half of the data. 
This analysis demonstrates that the data presented in
Tables~\ref{table:comp_od_HSF_TCF} and \ref{table:comp_ev_HSF_TCF} is
reliable.

\subsection{Structure of $\Theta_{od}^{3,TCF}$, $\Theta_{od}^{3,HSF}$ }

Table~\ref{table:comp_od_HSF_TCF} lists most of the elements of the
tensor $\Theta_{od}^{3,TCF}$ and $\Theta_{od}^{3,HSF}$.
The components are presented per class of elements, deduced from one another 
by elementary symmetries. 
The components not shown in Table~\ref{table:comp_od_HSF_TCF} can actually be 
derived from those shown by using incompressibility: $\partial_a u_a = 0$; 
they turn out to be small. 

Table~\ref{table:comp_od_HSF_TCF} shows
that the components of $\Theta_{od}^{3,TCF}$ vary relatively little in
the TCF, over
the range $ 256 \lesssim y^+ \lesssim 512$. This is consistent with 
the right panel of Fig.~\ref{fig:comp_TCF_HIT},
which shows that the 
norm $|| \Theta_{od}^{3,TCF} ||$ is essentially constant over this range 
(see the curve with the "x" symbols).

\begin{table}
\begin{center}
\begin{tabular}{| c | c | c | c | c | c | c |}
\cline{2-7}
\multicolumn{1}{c}{} & \multicolumn{6}{|c|}{  } \\
\multicolumn{1}{c}{} & \multicolumn{6}{|c|}{ Individual components of $ \Theta_{od}^{3,HSF}$, $\Theta_{od}^{3,TCF}$ ($\times 10$) } \\
\multicolumn{1}{c}{} & \multicolumn{6}{|c|}{  } \\
\hline
Moments & $\langle (\partial_2 u_1)^3 \rangle$ & $\langle (\partial_2 u_1)^2 (\partial_1 u_2) \rangle$ & $\langle \partial_2 u_1 (\partial_1 u_2)^2 \rangle $ & $\langle (\partial_1 u_2)^3  \rangle $ & & \\ 
HSF & $2.83$ & $-0.70$  & $0.40$ & $0.25$ &  &  \\
$y^+ = 256 $ & $5.98$ & $-1.31$ & $0.85$ & $-0.24$ &  &  \\
$y^+ = 512 $ & $5.57$ & $-1.29$ & $0.91$ & $-0.29$ &  &  \\ \hline \hline
Moments & $\langle \partial_2 u_1 (\partial_1 u_1)^2 \rangle$ & $\langle \partial_2 u_1 (\partial_2 u_2)^2 \rangle$ & $\langle \partial_2 u_1 (\partial_3 u_3)^2 \rangle $ & $\langle \partial_1 u_2 (\partial_1 u_1)^2 \rangle $ & $\langle \partial_1 u_2 (\partial_2 u_2)^2 \rangle $ & $\langle \partial_1 u_2 (\partial_3 u_3)^2 \rangle $\\ 
$HSF$ & $0.30$ & $0.24$ & $0.18$ & $-0.07$ & $-0.02$ & $0.04$  \\
$y^+= 256$ & $0.50$ & $0.48$ & $0.40$ & $-0.07$ & $-0.07$ & $0.03$  \\
$y^+= 512$ & $0.52$ & $0.50$ & $0.41$ & $-0.10$ & $-0.09$ & $0.02$  \\ \hline \hline
Moments & $\langle \partial_2 u_1 \partial_1 u_1 \partial_2 u_2 \rangle$ & $\langle \partial_2 u_1 \partial_2 u_2 \partial_3 u_3 \rangle$ & $\langle \partial_2 u_1 \partial_3 u_3 \partial_1 u_1 \rangle $ & $\langle \partial_1 u_2 \partial_1 u_1 \partial_2 u_2 \rangle $ & $\langle \partial_1 u_2 \partial_2 u_2 \partial_3 u_3 \rangle $ & $\langle \partial_1 u_2 \partial_3 u_3 \partial_1 u_1 \rangle $\\ 
$HSF$ & $-0.18$ & $-0.06$ & $-0.12$ & $0.06$ & $-0.04$ & $0.006$  \\
$y^+= 256$ & $-0.30$ & $-0.16$ & $-0.22$ & $0.10$ & $-0.02$ & $-0.01$  \\
$y^+= 512$ & $-0.31$ & $-0.18$ & $-0.22$ & $0.11$ & $-0.02$ & $-0.004$  \\ \hline \hline
Moments & $\langle \partial_2 u_1 (\partial_1 u_3)^2 \rangle$ & $\langle \partial_2 u_1 (\partial_3 u_1)^2 \rangle$ & $\langle \partial_2 u_1 \partial_1 u_3 \partial_3 u_1 \rangle $ & $\langle \partial_1 u_2 (\partial_1 u_3)^2 \rangle $ & $\langle (\partial_1 u_2) (\partial_3 u_1)^2 \rangle $ & $\langle (\partial_1 u_2) \partial_1 u_3 \partial_3 u_1 \rangle $\\ 
HSF & $0.33$ & $1.06$ & $-0.25$ & $-0.01$ & $-0.21$ & $0.03$  \\
$y^+= 256$ & $0.50$ & $1.84$ & $-0.41$ & $-0.01$ & $-0.38$ & $0.09$  \\
$y^+= 512$ & $0.60$ & $1.72$ & $-0.42$ & $-0.02$ & $-0.34$ & $0.09$ \\ \hline \hline
Moments & $\langle \partial_2 u_1 (\partial_2 u_3)^2 \rangle$ & $\langle \partial_2 u_1 (\partial_3 u_2)^2 \rangle$ & $\langle \partial_2 u_1 \partial_2 u_3 \partial_3 u_2 \rangle $ & $\langle \partial_1 u_2 (\partial_2 u_3)^2 \rangle $ & $\langle (\partial_1 u_2) (\partial_3 u_2)^2 \rangle $ & $\langle (\partial_1 u_2) \partial_2 u_3 \partial_3 u_2 \rangle $\\ 
HSF & $0.87$ & $-0.20$ & $-0.06$ & $-0.51$ & $0.19$ & $0.24$  \\
$y^+= 256$ & $1.51$ & $-0.11$ & $-0.08$ & $-0.86$ & $0.04$ & $0.34$  \\
$y^+= 512$ & $1.60$ & $0.03$ & $-0.15$ & $-0.83$ & $0.01$ & $0.36$  \\ \hline
Moments & $\langle \partial_3 u_2 \partial_2 u_2 \partial_1 u_3 \rangle$ & $\langle \partial_2 u_3 \partial_2 u_2 \partial_1 u_3 \rangle$ & $\langle \partial_3 u_2 \partial_2 u_2 \partial_3 u_1 \rangle $ & $\langle \partial_2 u_3 \partial_2 u_2 \partial_3 u_1 \rangle $ & & \\ 
HSF & $-0.18$ & $0.14$ & $0.32$ & $-0.29$ &  &  \\
$y^+= 256$ & $-0.30$ & $0.29$ & $0.48$ & $-0.48$ & &   \\
$y^+= 512$ & $-0.27$ & $0.24$ & $0.45$ & $-0.41$ &  &   \\ \hline
Moments & $\langle \partial_3 u_2 \partial_3 u_3 \partial_1 u_3 \rangle$ & $\langle \partial_2 u_3 \partial_3 u_3 \partial_1 u_3 \rangle$ & $\langle \partial_3 u_2 \partial_3 u_3 \partial_3 u_1 \rangle $ & $\langle \partial_2 u_3 \partial_3 u_3 \partial_3 u_1 \rangle $ & & \\ 
HSF & $0.17$ & $-0.18$ & $-0.25$ & $0.29$ &  &  \\
$y^+= 256$ & $0.24$ & $-0.33$ & $-0.37$ & $0.49$ & &   \\
$y^+= 512$ & $0.22$ & $-0.29$ & $-0.34$ & $0.40$ &  &   \\ \hline
\end{tabular}
\end{center}
\caption{A subset of the components of the
third order velocity gradient correlation tensor with an odd number of
indices $1$ and $2$ (all terms with an odd number of $3$ are zero by symmetry).
The values are shown for the HSF at $R_\lambda = 145$, and for the TCF 
at $y^+ = 256$ and $y^+ = 512$, which includes all the largest values measured.
With a few notable exceptions, these
values do not vary too much with $y^+$ in the range considered, and are 
roughly $\approx 1.8$ times larger than in the HSF.
}
\label{table:comp_od_HSF_TCF}
\end{table}
 
Among all the large terms listed in Table~\ref{table:comp_od_HSF_TCF}, the
largest values in HSF are smaller than the corresponding ones in TCF 
at $y^+ = 256$ 
and $y^+ = 512$ by a factor $\approx 1.8$. This is manifestly consistent with 
the value of
$\mu^{3}_{od}$ shown in Fig.~\ref{fig:comp_HSF_TCF}. The ratios between
the components in TCF and in HSF deviate 
much more significantly from $\approx 1.8$ for some of the weaker components.
Sign changes between HSF and TCF components are observed for the 
$\langle (\partial_1 u_2)^3 \rangle$
and $\langle \partial_2 u_1 (\partial_3 u_2)^2 \rangle$, which manifestly
point to differences between the two flows.

 The largest of all the components shown in Table~\ref{table:comp_od_HSF_TCF}
is $\langle (\partial_2 u_1)^3 \rangle$, which exceeds any of the other
components by a factor $ \approx 3$. We recall that
the term $\partial_2 u_1$
is the fluctuation around the mean shear, $dU_1/dx_2$, and that the third
moment 
$\langle (\partial_2 u_1)^3 \rangle$ has been used
to investigate anisotropy in turbulent shear 
flows~\cite{PumShr95,Pumir96,Pumir+16}.  \\
Particularly significant is the sign of 
$\langle (\partial_2 u_1)^3 \rangle$, which is identical to the sign 
of the mean shear, $dU_1/dx_2$. This reflects the
(partial) expulsion of the velocity gradients from large regions of the 
flow~\cite{PumShr95,Pumir+16}.
In the related problem of a passive scalar, $\theta$,
in the presence of a mean scalar
gradient, $\langle \theta \rangle = \mathbf{G} \cdot \mathbf{x}$,
it is well-established that the distribution of the scalar gradient,
$\nabla \theta$, has a sharp peak at $\nabla \theta = 0$. This points to a
sharp expulsion of the gradients, which are manifested by the presence of 
large regions of space over which the scalar is approximately constant. 
These regions are separated by narrow regions, where large scalar jumps
form~\cite{Sreeni91,HolzSig94,Pumir94}, implying the formation of large
gradients, which in turn contribute to 
the odd moments of 
$\mathbf{G} \cdot (\nabla \theta)$, in particular to the skewness, which
is determiend by the large scale gradient, $\mathbf{G}$.
This effect is also seen, although in a weaker form,
in numerical simulations of HSF or TCF. In the flows considered here, 
the mean velocity gradient, $dU_1/dx_2$ is always positive, and
the positive sign of $\langle (\partial_2 u_1) ^3 \rangle$ can be interpreted
as a result of very large positive fluctuations of $\partial_2 u_1$, 
resulting from
extended regions of space where $\partial_2 (U_1 + u_1)$ is relatively small, 
separated
by regions where $\partial_2 (U_1 + u_1)$ is much larger than the 
mean, $\partial_2 U_1$. \\

The picture sketched above would suggest that
quantities of the form 
$\langle \partial_2 u_1 (\partial_a u_b \partial_a u_b) \rangle$,
could be dominated by positive fluctuations of 
$\partial_2 u_1$, suggesting a positive value of the moment
$\langle \partial_2 u_1 (\partial_a u_b)^2 \rangle$.
This is true for all terms with
$a = b$, and for all terms with $a \ne b$, with the exception
of the
term $\langle \partial_2 u_1 (\partial_3 u_2)^2 \rangle$ in HSF. 
This term turns out to be relatively small in TCF, and
even to becomes positive in TCF at $y^+ = 512$. 
The same considerations suggest that terms of the form 
$\langle \partial_2 u_1 (\partial_a u_b \partial_b u_a) \rangle$ have the
same sign as $\langle (\partial_a u_b \partial_b u_a) \rangle$, which
is negative when $a \ne b$. This is consistent with the expressions listed in
Table~\ref{table:comp_od_HSF_TCF}.
Last, we also notice that terms of the form 
$\langle \partial_2 u_1 (\partial_a u_a \partial_b u_b ) \rangle$ for $ a \ne b$
all have negative signs, as implied by the above considerations. Obtaining
a more quantitative parametrization of terms of the form 
$\langle \partial_2 u_1 (\partial_a u_b \partial_c u_d) \rangle$
in terms of $\langle \partial_a u_b \partial_c u_d \rangle$ does not appear
to be simply feasible.

Whereas the terms containing one or more terms of the form $\partial_2 u_1$
are the dominant ones, and have a sign that can be understood with the 
help of the simple considerations above, the third moments 
containing $\partial_1 u_2$ are generally smaller than those with 
$\partial_2 u_1$.
In TCF, the first set of components in 
Table~\ref{table:comp_od_HSF_TCF} indicates that the sign
of $\langle (\partial_2 u_1)^{3-p} (\partial_1 u_2)^p \rangle$ is $(-1)^p$,
generally indicating a sign difference in the contributions of 
$\partial_2 u_1$ (which tend to be positive) and of $\partial_1 u_2 $ 
(which tend to be negative).
This empirical rule accounts for the sign of most of the terms of the
form $\langle \partial_1 u_2 (\partial_a u_b)^2 \rangle$ in the table,
with a few exceptions.
The difference in sign of 
$\langle (\partial_1 u_2)^3 \rangle$ between TCF and HSF points to the
quantitative limitation
of the empirical observation that $(\partial_1 u_2)$ generally provides
a negative contribution to the moments investigated here.

\subsection{Structure of $\Theta_{ev}^{3,TCF}$, $\Theta_{ev}^{3,HSF}$ }

\begin{table}
\begin{center}
\begin{tabular}{| c | c | c | c | c | c | c |}
\cline{2-7}
\multicolumn{1}{c}{} & \multicolumn{6}{|c|}{  } \\
\multicolumn{1}{c}{} & \multicolumn{6}{|c|}{ Individual components of $ \Theta_{ev}^{3,HSF}$, $\Theta_{ev}^{3,TCF}$ ($\times 10$) } \\
\multicolumn{1}{c}{} & \multicolumn{6}{|c|}{  } \\
\hline
 Moments & $\langle (\partial_1 u_1)^3 \rangle$ & $\langle (\partial_2 u_2)^3 \rangle$ & $\langle (\partial_3 u_3)^3 \rangle$ & & & \\ 
 HSF & $-0.18$ & $0.06$ & $0.03$ & & & \\ 
 $y^+=256$ & $-0.37$ & $0.18$ & $-0.007$ & & & \\ 
 $y^+=512$ & $-0.28$ & $0.15$ & $-0.02$ & & & \\ \hline \hline
 Moments & $\langle (\partial_1 u_1) (\partial_1 u_2)^2 \rangle$ & $\langle (\partial_1 u_1) (\partial_1 u_3)^2\rangle$ & $\langle (\partial_2 u_2) (\partial_2 u_1)^2 \rangle$ & $ \langle (\partial_2 u_2)(\partial_2 u_3)^2 \rangle$ & $\langle (\partial_3 u_3) (\partial_3 u_1)^2 \rangle$ & $\langle (\partial_3 u_3) (\partial_3 u_2)^2 \rangle$ \\ 
HSF & $-0.26$ & $-0.27 $ & $0.22$ & $0.11 $ & $0.14$ & $-0.06$ \\ 
$y^+=256$ & $-0.44$ & $-0.49 $ & $0.58$ & $0.41 $ & $0.21$ & $-0.21$ \\ 
$y^+=512$ & $-0.38$ & $-0.40 $ & $0.48$ & $0.31 $ & $0.15$ & $-0.17$ \\ \hline \hline
 Moments & $\langle (\partial_1 u_1) (\partial_2 u_1)^2 \rangle$ & $\langle (\partial_1 u_1) (\partial_3 u_1)^2\rangle$ & $\langle (\partial_2 u_2) (\partial_1 u_2)^2 \rangle$ & $ \langle (\partial_2 u_2)(\partial_3 u_2)^2 \rangle$ & $\langle (\partial_3 u_3) (\partial_1 u_3)^2 \rangle$ & $\langle (\partial_3 u_3) (\partial_2 u_3)^2 \rangle$ \\ 
HSF & $-0.14$ & $-0.14 $ & $0.13$ & $0.12 $ & $0.05$ & $0.01$ \\ 
$y^+=256$ & $-0.12$ & $-0.25 $ & $0.11$ & $0.27 $ & $-0.05$ & $-0.05$ \\ 
$y^+=512$ & $-0.10$ & $-0.18 $ & $0.17$ & $0.23 $ & $-0.04$ & $-0.08$ \\ \hline \hline
 Moments & $\langle \partial_2 u_1 \partial_1 u_1 \partial_1 u_2 \rangle$ & $\langle \partial_3 u_1 \partial_1 u_1 \partial_1 u_3\rangle$ & $\langle \partial_1 u_2 \partial_2 u_2 \partial_2 u_1 \rangle$ & $ \langle \partial_3 u_2 \partial_2 u_2 \partial_2 u_3 \rangle$ & $\langle \partial_1 u_3 \partial_3 u_3 \partial_3 u_1 \rangle$ & $\langle \partial_2 u_3 \partial_3 u_3 \partial_3 u_2 \rangle$ \\ 
HSF & $0.14$ & $0.14 $ & $-0.13$ & $-0.08 $ & $-0.06$ & $0.03$ \\ 
$y^+=256$ & $0.26$ & $0.24 $ & $-0.24$ & $-0.26 $ & $-0.07$ & $0.13$ \\ 
$y^+=512$ & $0.23$ & $0.19 $ & $-0.20$ & $-0.20 $ & $0.06$ & $0.10$ \\ \hline \hline
 Moments & $\langle \partial_1 u_1 (\partial_2 u_3)^2 \rangle$ & $\langle \partial_1 u_1 (\partial_3 u_2)^2 \rangle$ & $\langle \partial_2 u_2 (\partial_1 u_3)^2 \rangle$ & $ \langle \partial_2 u_2 ( \partial_3 u_1)^2 \rangle$ & $\langle \partial_3 u_3 (\partial_1 u_2)^2  \rangle$ & $\langle \partial_3 u_3 ( \partial_2 u_1 )^2 \rangle$ \\ 
HSF & $-0.12$ & $-0.08 $ & $0.21$ & $-0.08 $ & $0.13$ & $-0.09$ \\ 
$y^+=256$ & $-0.33$ & $-0.36 $ & $0.55$ & $0.02 $ & $0.35$ & $-0.48$ \\ 
$y^+=512$ & $-0.21$ & $-0.04 $ & $0.45$ & $0.02 $ & $0.26$ & $-0.40$ \\ \hline 
\hline
 Moments & $\langle \partial_1 u_2 \partial_2 u_3 \partial_1 u_3 \rangle$ & $\langle \partial_1 u_2 \partial_3 u_2 \partial_3 u_1 \rangle$ & $\langle \partial_1 u_2 \partial_3 u_2 \partial_1 u_3 \rangle$ & $ \langle \partial_2 u_1  \partial_1 u_3 \partial_2 u_3 \rangle$ & $\langle \partial_2 u_1 \partial_3 u_1 \partial_3 u_2  \rangle$ & $\langle \partial_2 u_1 \partial_3 u_1 \partial_2 u_3  \rangle$ \\ 
HSF & $0.01$ & $-0.02 $ & $-0.08$ & $-0.08 $ & $0.10$ & $0.19$ \\ 
$y^+=256$ & $-0.10$ & $-0.09 $ & $-0.30$ & $-0.02 $ & $0.27$ & $0.40$ \\ 
$y^+=512$ & $-0.05$ & $-0.08 $ & $-0.23$ & $-0.02 $ & $0.25$ & $0.23$ \\ \hline
\end{tabular}
\end{center}
\caption{A subset of the components of the deviations 
of the third order velocity gradient correlation tensor with an even
number of all indices, from the corresponding isotropic tensor.
The values are shown for the HSF at $R_\lambda = 145$, and for the TCF 
at $y^+ = 256$ and $y^+ = 512$. As it was the case for the tensors 
$\Theta_{od}^{3,HSF} $ and $\Theta_{od}^{3,TCF}$, see 
Table.~\ref{table:comp_od_HSF_TCF}. The values for the TCF
vary more with $y^+$ than in the case of the odd terms, consistent with
Fig.~\ref{fig:comp_TCF_HIT}. The ratio between the values in TCF and in HSF is
roughly $ 1.8$.
}
\label{table:comp_ev_HSF_TCF}
\end{table}

Table \ref{table:comp_ev_HSF_TCF} lists several groups of coefficients of 
$\Theta_{ev}^{3,HSF}$ and $\Theta_{ev}^{3,TCF}$, which include all the
largest elements of $\Theta_{ev}^{3,HSF}$. Namely, we separated all the 
non-zero elements of $\Theta_{ev}^{3,HIT}$ in several classes of identical 
elements,
which can be deduced from one another by a permutation of the indices 
$(1, 2, 3)$ (see the Appendix), and selected those for which one of the 
elements of this class is larger than $1/100$.
Table \ref{table:comp_ev_HSF_TCF} systematically compares the results obtained
for HSF with those obtained for TCF, for $y^+ = 256$ and $y^+ = 512$.
The values of $\Theta_{ev}^{3,TCF}$ generally show
a larger variation over the range $256 \lesssim y^+ \lesssim 512$,
compared to the variation found for the components of $\Theta_{od}^{3,TCF}$
shown in Table~\ref{table:comp_od_HSF_TCF}. This is consistent with 
Fig.~\ref{fig:comp_TCF_HIT}, which shows a small, but visible variation of
$||\Theta_{ev}^{3,TCF}||$ over the corresponding range of values of $y^+$.

Overall, Table~\ref{table:comp_ev_HSF_TCF} indicates that, consistent with
the value of $\mu_{ev}^3$ reported in Fig.~\ref{fig:comp_HSF_TCF}, the ratios
between many of the coefficients shown in the table are close to
$\approx 1.8$. This is particularly true for the largest elements of the
tensor.
This proportionality relation does not hold so well for some
of the smaller coefficients. 
Some of the coefficients
of $\Theta_{ev}^{3,HSF}$ differ from those of $\Theta_{ev}^{3,TCF}$ even
by their signs. The differences between the components 
of $\Theta_{ev}^{3,TCF}$ and $\approx \mu_{ev} \Theta_{ev}^{3,HSF}$, 
tends to be systematically larger than their counterparts for the odd 
components of $\Theta$. This
certainly accounts for the fact that $||\Psi_{ev}||$ is roughly
twice larger than $|| \Psi_{od}||$, see the lower right panel of
Fig.~\ref{fig:comp_HSF_TCF}.

\section{Discussion}
\label{sec:discussion}

To recapitulate, we have systematically investigated the structure
of the second and third order correlation of the velocity gradient tensor
in elementary shear flows: HSF and TCF in the logarithmic region. 
The comparison with the exact form of the tensors in the case of HIT
provides an unambiguous way to measure how the
flow deviates from isotropy. 
In particular, our analysis allows us to identify corrections to the HIT
form of the tensor. Although the magnitude of these corrections depends 
on the flow (TCF or HSF), the overall structures are very comparable. 
It is worth stressing here that the largest discrepancy between the measured
velocity tensors, and its HIT counterpart comes from the odd contribution
to the $n = 3$ velocity gradient correlations (see the right panels of
Fig.~\ref{fig:comp_TCF_HIT}). The corresponding structure is the one that
turns out to be the most similar, in TCF and in HSF (see the lower right 
panel of Fig.~\ref{fig:comp_HSF_TCF}).

Interestingly, the largest of all the components of the odd contribution to 
the third order velocity derivative 
tensor is the one that corresponds to the third order moment of 
$\partial_2 u_1$, the derivative of the streamwise component of the velocity
in the normal direction. This quantity had been investigated 
numerically~\cite{PumShr95,Pumir96,Schum00,Pumir+16} 
and experimentally~\cite{GargWarh98,ShenWarh00}, based on a possible analogy
with the 
observation of a skewness of the scalar gradient, which remains of 
order ${\cal O}(1)$, independent of $R_\lambda$, in the presence
of a mean scalar gradient~\cite{Sreeni91,HolzSig94,Pumir94,TongWarh94,ShrSig00,Warhaft00}. While the scalar analogy was providing 
an enticing theoretical motivation, suggesting experimentally measurable 
quantities, the results of the present study demonstrate that the third 
moment of $\partial_2 u_1$ is indeed the most sensitive third order moment
to the presence of shear.
Consistent with the ratio of an overall factor $\approx 1.7$ between
the odd part of the third order velocity gradient tensor in the TCF 
for $200 \lesssim y^+ \lesssim 600$ and HSF, the skewness of $\partial_2 u_1$
was found to $\approx 1.07$ in TCF, and $0.65$ in HSF~\cite{Pumir+16}.
 
As it turns out, the Reynolds numbers in the log-layer of the TCF and in the HSF
are comparable, $R_\lambda \approx 150$. 
Comparing TCF and HSF at comparable Reynolds numbers leads already to 
interesting conclusions. It has already been observed that the
difference in the magnitude of the third moment of the velocity gradient
tensors is presumably due to the very different forcing. As suggested by
standard phenomenology, the notion that the flow integral scale in TCF a 
distance $x_2$ away from the wall is $\approx x_2$~\cite{Lumley,Pope} 
suggests that the effect of the boundary may be felt throughout the log-region
of the TCF. On the other hand, the skewness of $\partial_2 u_1$ in HSF is known
to decrease, albeit weakly, with $R_\lambda$ (like $R_\lambda^{-\alpha}$, with
$\alpha \approx 0.5$). 
For obvious reasons, it would be interesting to study the dependence of
the low-order velocity gradient tensors considered here as a function of
the Reynolds number.
Although the even tensor $\Theta_{ev}^{3,HSF}$ and $\Theta_{ev}^{3,TCF}$
are generally smaller than their odd counterparts in the Reynolds numbers 
considered here, it remains to be understood whether and how these terms 
decay as a function of the Reynolds number. The limited comparison between 
$||\Theta_{ev}^{3,HSF} || $ at the two Reynolds numbers may suggest a 
faster decay of $|| \Theta_{ev}^{3,HSF}|| $ than the one observed for
$|| \Theta_{od}^{3,HSF}||$. Convincing evidence can only be provided
by investigating HSF at higher Reynolds numbers.   
Although it is to be expected that a systematic group theoretic
analysis of the problem may be helpful~\cite{BifProc05}, especially
in the case of a homogeneous flow, a precise
calculation remains to be done. The decomposition
of the tensors $\Theta^{n,flow}$ as a sum of an even and of an odd
part merely separates spherical harmonics with even and
odd eigenvalues $l$ of the angular momentum operator. 
It is generally expected that the decay of the tensors
will be dominated by the lowest angular momentum ($l = 1$ for the odd terms, 
and $l = 2$ for the even ones).
In the inhomogeneous case of TCF, where the role of the boundary may persist
even far away from the boundary, whether the skewness will stay constant, or 
ultimately decay as in the HSF case, remains to be investigated.

 It is enticing to compare the structure of the velocity gradient 
correlation tensor, uncovered in the case of HSF
and TCF in the logarithmic layer, with the structure obtained by shearing a
homogeneous isotropic flow. In practice, this can be done by applying 
the transformation induced by the mean shear~\cite{Townsend70} to a numerical
solution of the Navier-Stokes equations, and deducing the velocity gradient
correlation tensors. This can be done in practice by using 
a simulation in a triply periodic domain (see~\cite{Pumir94} for details)
and applying a stretching in the range $0 \le S t \le 1$. 
The deviations $\Theta$ of the second and third order
velocity gradient correlation tensors from the form expected in a
homogeneous, isotropic flow,
defined using Eq.~\eqref{eq:decomp_T}, grows linearly with 
the amount of stretching. 
The structure of $\Theta$, however, does not indicate any close similarity 
with the one found in HSF and TCF. 

The observation that the tensors of order 
$n = 2$ and $n = 3$ in HIT are completely characacterized by only one
invariant makes our analysis unambiguous. Higher order tensors in HIT
involve more than 1 dimensional quantity, which may significantly 
complicate the analysis. As recently observed~\cite{Pumir+16}, however,
in the case $n = 4$, the ratios between the 4 invariants quantities
that characterize the velocity gradient correlations~\cite{Siggia81} seem to be
independent of the flow. This property, which remains to be more thoroughly 
understood, may make the analysis tractable also for the $n=4$ order tensor.
Such an analysis may provide new insight on the production and formation
of very large gradients in the flows. The available experimental data in 
the case $n > 4$~\cite{ShenWarh02} points to a rich structure, which 
deserves further attention.

\medskip
\section*{Acknowledgements}

This work originated from discussions with Eric Siggia. 
It is a pleasure to thank him, along with Haitao Xu, for many useful remarks.
Some of the computations were performed at the PSMN computing centre at 
the Ecole Normale Sup\'erieure de Lyon.

\section*{Appendix}

As done in~\cite{Siggia81}, it is convenient to decompose the velocity gradient tensor $A_{ab} 
= \partial_a u_b$, where $1 \le a \le 3$ and $1 \le b \le 3$ refer to
the spatial coordinates, into a symmetric and antisymmetric parts:
$s_{ab} = (A_{ab} + A_{ba})/2$, and $\omega_a = \epsilon_{abc} A_{bc}$,
so 
\begin{equation}
\partial_a u_b = s_{ab} + \epsilon_{abc} \omega_c/2
\label{eq:expr_dv}
\end{equation}

The third moment $\langle s_{ab} s_{cd} s_{ef} \rangle$ can be expressed
as a function of a single dimensional quantity, $\langle tr(s^3) \rangle$, 
and of the (Kronecker) $\delta$-tensor.
Using the isotropy of the tensor $s$, one can write, in full generality:
\begin{eqnarray}
\langle s_{ab} s_{cd} s_{ef} \rangle & = & A \delta_{ab} \delta_{cd} \delta_{ef} \nonumber \\
&+ & B \Bigl( 
\delta_{ab} ( \delta_{ce} \delta_{df} + \delta_{cf} \delta_{de} ) +
\delta_{cd} ( \delta_{ae} \delta_{bf} + \delta_{af} \delta_{be} ) +
\delta_{ef} ( \delta_{ac} \delta_{bd} + \delta_{ad} \delta_{bc} )  \Bigr) \nonumber \\
& + & C \Bigl(
\delta_{ac} \delta_{de} \delta_{fb} + \delta_{ac} \delta_{df} \delta_{eb} +
\delta_{ad} \delta_{ce} \delta_{fb} + \delta_{ad} \delta_{cf} \delta_{eb} \nonumber \\
& & ~~ +\delta_{bc} \delta_{de} \delta_{fa} + \delta_{bc} \delta_{df} \delta_{ea} +
\delta_{bd} \delta_{ce} \delta_{fa} + \delta_{bd} \delta_{cf} \delta_{ea} \Bigr)
\label{eq:s3_general}
\end{eqnarray}

Constraints on the constants $A$, $B$ and $C$ can be found by expressing the
incompressibility of the velocity field: $\sum_{a=1}^3 s_{aa} = 0$:
\begin{eqnarray}
\sum_{a=1}^3 \langle s_{aa} s_{cd} s_{ef} \rangle & = & 3 A \delta_{cd} \delta_{ef} 
+  B \Bigl( 
3 ( \delta_{ce} \delta_{df} + \delta_{cf} \delta_{de} ) +
4 \delta_{cd} \delta_{ef} \Bigr) 
 +  4 C \Bigl(
\delta_{de} \delta_{fc} + \delta_{ce} \delta_{df} \Bigr) \\ \nonumber
& = & (3A + 4B) \delta_{cd} \delta_{ef} + (3 B + 4 C) ( \delta_{de} \delta_{cf} + \delta_{ce} \delta_{df} )
\label{eq:s3_incomp}
\end{eqnarray}
which immediately imposes that $3A + 4B = 3B + 4C = 0$, and thus allows us to
express $\langle s_{ab} s_{cd} s_{ef} \rangle$ with only one number. 
To fully determine the tensor, we evaluate the trace of $s^3$: 
$\langle tr(s^3) \rangle = \sum_{a,b,c} \langle s_{ab} s_{bc} s_{ca} \rangle$
\begin{equation}
\langle tr (s^3 ) \rangle = 3 A + 36 B + 66 C
\label{eq:trs3}
\end{equation}
which leads to the final expression:
\begin{eqnarray}
\langle s_{ab} s_{cd} s_{ef} \rangle & = & \frac{2}{35} \langle tr(s^3) \rangle \times \Bigl[ \frac{4}{3} \delta_{ab} \delta_{cd} \delta_{ef} \nonumber \\
&- & [ 
\delta_{ab} ( \delta_{ce} \delta_{df} + \delta_{cf} \delta_{de} ) +
\delta_{cd} ( \delta_{ae} \delta_{bf} + \delta_{af} \delta_{be} ) +
\delta_{ef} ( \delta_{ac} \delta_{bd} + \delta_{ad} \delta_{bc} )  ] \nonumber \\
& + & \frac{3}{4} (
\delta_{ac} \delta_{de} \delta_{fb} + \delta_{ac} \delta_{df} \delta_{eb} +
\delta_{ad} \delta_{ce} \delta_{fb} + \delta_{ad} \delta_{cf} \delta_{eb} \nonumber \\
& & ~~ +\delta_{bc} \delta_{de} \delta_{fa} + \delta_{bc} \delta_{df} \delta_{ea} +
\delta_{bd} \delta_{ce} \delta_{fa} + \delta_{bd} \delta_{cf} \delta_{ea} ) \Bigr]
\label{eq:s3_final}
\end{eqnarray}

In addition to the tensor $\langle s_{ab} s_{cd} s_{ef} \rangle$, one has also
to consider quantities such as 
$\langle s_{ab} s_{cd} \omega_e \rangle$,
$ \langle s_{ab} \omega_c \omega_d \rangle$ and 
$\langle \omega_a \omega_b \omega_c \rangle$.
Under reflection, 
$(x, y , z) \rightarrow (-x, -y, -z) $, the tensor $s$ remains unchanged, 
whereas $\omega \rightarrow -\omega$.
Thus, in a flow that is invariant under reflection, only terms with an 
even number of values of $\omega_a$ are non zero. The only term to estimate
is therefore $\langle s_{ab} \omega_c \omega_d \rangle$.
As before, symmetry considerations impose that:
\begin{equation}
\langle s_{ab} \omega_c \omega_d \rangle = D \delta_{ab} \delta_{cd} 
+ E ( \delta_{ac} \delta_{bd} + \delta_{ad} \delta_{bc}  )\label{eq:somom_gen}
\end{equation}
Incompressibility ($\sum_{1 \le a \le 3} s_{aa} = 0$) leads to:
$3D + 2E = 0$.
The tensor can be conveniently expressed in terms of stretching,
$\langle \mathbf{\omega} \cdot \mathbf{s} \cdot \mathbf{\omega} \rangle$:
\begin{equation}
\langle s_{ab} \omega_c \omega_d \rangle = 
\langle \mathbf{\omega} \cdot \mathbf{s} \cdot \mathbf{\omega} \rangle
\Bigl( - \frac{1}{15} \delta_{ab} \delta_{cd} + \frac{1}{10} ( \delta_{ac} \delta_{bd} + \delta_{ad} \delta_{bc} )
\Bigr)
\label{eq:somom_fin}
\end{equation}

Remains to express the full tensor 
$\langle \partial_a u_b \partial_c u_d \partial_e u_f \rangle$. 
This can be done by using Eq.~(\ref{eq:expr_dv}):
\begin{eqnarray}
\langle \partial_a u_b \partial_c u_d \partial_e u_f \rangle
& = & \langle 
(s_{ab} + \epsilon_{abA} \omega_A/2 )
(s_{cd} + \epsilon_{cdB} \omega_B/2 )
(s_{ef} + \epsilon_{efC} \omega_C/2 ) \rangle \nonumber \\
& = & \langle s_{ab} s_{cd} s_{ef} \rangle +
\frac{1}{4} \epsilon_{cdB} \epsilon_{efC} \langle s_{ab} \omega_B \omega_C \rangle \nonumber \\
& + &\frac{1}{4} \epsilon_{efC} \epsilon_{abA} \langle s_{cd} \omega_C \omega_A \rangle +
\frac{1}{4} \epsilon_{abA} \epsilon_{cdB} \langle s_{ef} \omega_A \omega_B \rangle 
\label{eq:dvdvdv_exp}
\end{eqnarray}

By substituting the expressions for $\langle s_{ab} s_{cd} s_{ef} \rangle$ and
$\langle s_{ab} \omega_A \omega_B \rangle$, Eq.~(\ref{eq:s3_final},\ref{eq:somom_fin}) into
Eq.~(\ref{eq:dvdvdv_exp}) gives:
\begin{eqnarray}
& & \langle \partial_a u_b \partial_c u_d \partial_e u_f \rangle 
= \frac{2}{35} \langle tr(s^3) \rangle \times \Bigl[ \frac{4}{3} \delta_{ab} \delta_{cd} \delta_{ef} \nonumber \\
&- & [ 
\delta_{ab} ( \delta_{ce} \delta_{df} + \delta_{cf} \delta_{de} ) +
\delta_{cd} ( \delta_{ae} \delta_{bf} + \delta_{af} \delta_{be} ) +
\delta_{ef} ( \delta_{ac} \delta_{bd} + \delta_{ad} \delta_{bc} )  ] \nonumber \\
& + & \frac{3}{4} (
\delta_{ac} \delta_{de} \delta_{fb} + \delta_{ac} \delta_{df} \delta_{eb} +
\delta_{ad} \delta_{ce} \delta_{fb} + \delta_{ad} \delta_{cf} \delta_{eb} \nonumber \\
& & ~~ +\delta_{bc} \delta_{de} \delta_{fa} + \delta_{bc} \delta_{df} \delta_{ea} +
\delta_{bd} \delta_{ce} \delta_{fa} + \delta_{bd} \delta_{cf} \delta_{ea} ) \Bigr] \nonumber \\
& + & \frac{1}{20} \langle \mathbf{\omega} \cdot \mathbf{s} \cdot \mathbf{\omega} \rangle \times \Bigl[ - \frac{1}{3} \delta_{ef} ( \delta_{ac} \delta_{bd} - \delta_{ad} \delta_{bc} ) - \frac{1}{3}
\delta_{cd} ( \delta_{ae} \delta_{bf} - \delta_{af} \delta_{be} ) - \frac{1}{3}
\delta_{ab} ( \delta_{ce} \delta_{df} - \delta_{cf} \delta_{de} ) \nonumber \\
& + & \frac{1}{2} (\epsilon_{abe} \epsilon_{cdf} + \epsilon_{abf} \epsilon_{cde} ) 
+ \frac{1}{2} ( \epsilon_{abc} \epsilon_{efd} + \epsilon_{abd} \epsilon_{efc} ) 
+ \frac{1}{2} ( \epsilon_{cda} \epsilon_{efb} + \epsilon_{cdb} \epsilon_{efa} ) \Bigr] 
\label{eq:dvdvdv_int}
\end{eqnarray}

The two constants involved in the expression Eq.~(\ref{eq:dvdvdv_int}) are in 
fact connected by the well known identity:
\begin{equation}
\sum_{a,b,c = 1}^3 \langle \partial_a v_b \partial_b v_c \partial_c v_a \rangle = 0
\label{eq:identity_Betchov} 
\end{equation}
which results from the homogeneity of the flow~\cite{Betchov56}.
Elementary algebraic manipulations lead to:
\begin{equation}
\sum_{a,b,c = 1}^3 \langle \partial_a v_b \partial_b v_c \partial_c v_a \rangle = \langle tr(s^3) \rangle + \frac{3}{4} \langle \mathbf{\omega} \cdot \mathbf{s} \cdot \mathbf{\omega} \rangle
\label{eq:interm}
\end{equation}
hence to the well known identity: 
\begin{equation}
\langle \mathbf{\omega} \cdot \mathbf{s} \cdot \mathbf{\omega} \rangle = - \frac{4}{3} \langle tr(s^3) \rangle
\label{eq:rel_betch}
\end{equation}

Substituting Eq.~\eqref{eq:rel_betch} into Eq.~\eqref{eq:dvdvdv_int} immediately
leads to Eq.~\eqref{eq:dvdvdv_hit}.

It is very elementary to see that only components of $T^{3,HIT}_{abcdef}$ with 
an even number of indices $1$, $2$ and $3$ can be non-zero.
The components of $T^{3,HIT}$ which involve only one of the three indices 
are $T^{3,HIT}_{aaaaaa} = \frac{8}{105}$. Those which involve two indices,
$a$ and $b$ ($a \ne b$) are one of the possible forms:
$T^{3,HIT}_{aabbbb} = -\frac{4}{105}$, $T^{3,HIT}_{aaabab} = \frac{16}{315}$, 
$T^{3,HIT}_{abbaba} = \frac{16}{315}$ and 
$T^{3,HIT}_{aaabba} = \frac{2}{315}$.
Last, the components of $T^{3,HIT}$ which involve 3 diffent indices,
$a$, $b$ and $c$, with $a \ne b$, $b \ne c$ and $a \ne c$ are of the form:
$T^{3,HIT}_{aabbcc} = \frac{8}{105}$, 
$T^{3,HIT}_{aabcbc} = -\frac{32}{315}$, 
$T^{3,HIT}_{aabccb} = -\frac{4}{315}$, 
$T^{3,HIT}_{abbcca} = -\frac{2}{35}$ and 
$T^{3,HIT}_{abbcac} = -\frac{8}{105}$.

%\bibliography{references}
%merlin.mbs apsrev4-1.bst 2010-07-25 4.21a (PWD, AO, DPC) hacked
%Control: key (0)
%Control: author (0) dotless jnrlst
%Control: editor formatted (1) identically to author
%Control: production of article title (0) allowed
%Control: page (1) range
%Control: year (0) verbatim
%Control: production of eprint (0) enabled
\providecommand{\noopsort}[1]{}

\end{document}